\documentclass{aa}  
\usepackage{graphicx}
\usepackage{natbib}
\bibpunct{(}{)}{;}{a}{}{,}  
\usepackage[varg]{txfonts}
\usepackage{epstopdf}
\begin{document} 

   \title{Abundance analysis of a CEMP-no star in the Carina dwarf spheroidal galaxy\thanks{Based on observations collected at the European Southern Observatory at Paranal, Chile; Large Programme proposal 171.B- 0520.} 
\fnmsep\thanks{Table A.1 is  available in electronic form
at the CDS via anonymous ftp to cdsarc.u-strasbg.fr (130.79.128.5)
or via http://cdsweb.u-strasbg.fr/cgi-bin/qcat?J/A+A/}
}
   \author{
   A. Susmitha\inst{1,2}
   \and
   A. Koch\inst{3}
   \and 
   T. Sivarani\inst{1}
   }
\authorrunning{A. Susmitha, A. Koch, \& T. Sivarani}
\titlerunning{CEMP-no star in the Carina dwarf spheroidal}
\offprints{A. Susmitha;  \email{susmitha@iiap.res.in; a.koch1@lancaster.ac.uk; sivarani@iiap.res.in}}
   \institute{Indian Institute of Astrophysics, Bangalore-34, India
         \and
             Indian Institute of Science, Bangalore-12, India
         \and
             Department of Physics, Lancaster University, Lancaster, LA1, 4YB, United Kingdom 
             }
\date{} 
\abstract
{Carbon-enhanced metal-poor (CEMP) stars bear important imprints of the early chemical enrichment of any stellar system.
While these stars are known to exist in copious amounts in the Milky Way halo, detailed chemical abundance 
data from the faint dwarf spheroidal (dSph) satellites are still sparse, although the relative fraction of these stars increases with decreasing metallicity.
Here, we report the abundance analysis of a metal-poor ([Fe/H]=$-2.5$ dex), carbon-rich ([C/Fe]=1.4 dex) 
star, ALW-8, in the Carina dSph using high-resolution spectroscopy obtained with the ESO/UVES instrument. 
Its spectrum does not indicate any over-enhancements of neutron 
capture elements. Thus classified as a CEMP-no star, this is the first detection of this kind of star in Carina. 
Another of our sample stars, ALW-1, is shown to be a CEMP-$s$ star, but its immediate binarity prompted us to discard it from a detailed analysis.
The majority of the 18 chemical elements we measured are typical of Carina's field star population and 
also agree with CEMP stars in other dSph galaxies. 
Similar to the only known CEMP-no star in the Sculptor dSph and the weak-$r$-process star HD 122563, the lack of any strong barium-enhancement is
accompanied by a moderate overabundance in yttrium, indicating a weak $r$-process activity. 
The overall abundance pattern confirms that, also in Carina,  the formation site for CEMP-no stars has been affected by both faint supernovae and by  standard core collapse supernovae. 
Whichever process was responsible for the heavy element production in ALW-8 must be a ubiquitous source to pollute the CEMP-no stars,  
acting independently of the environment such as in the Galactic halo or in dSphs. 
}
   \maketitle
\section{Introduction}
Observational studies of metal-poor stars in the  Milky Way (MW) and other nearby galaxies reveal that the early Universe has experienced  various nucleosynthetic mechanisms. The diverse chemical composition these stars preserve in their photosphere is the result of such nucleosynthetic mechanisms and detailed studies of 
their abundances yield information about the nature of their progenitors and how the latter contributed to the interstellar medium (ISM) through various (explosive)  mechanisms. The diversity in the stellar  chemical properties becomes more important and unique as the  metallicity of the stars decreases. 

Metal-poor stars in the MW have been   extensively studied and classified into different groups based on their metallicity, 
with further distinctions based on the abundance of certain key elements \citep[e.g.,][]{beers-christlieb2005ARA&A,aoki2007apj}.
Amongst these, Carbon-enhanced metal-poor (CEMP) stars (defined by [Fe/H] $< -2.0$ and [C/Fe] $>$0.7) 
are of great interest since the origin of the carbon overabundance is versatile and closely related to the formation and enrichment history of 
the host stellar system \citep{aoki2007apj,masseron2010aa,Carollo2014,bonifacio2015aa,cjhansen2016,Koch2016,Lee2017}.  
These stars are primarily categorized  based on the presence and absence of neutron capture elements as 
CEMP-$s$, CEMP-$r$, or  CEMP-$r/s$, versus CEMP-no, respectively.  
Amongst these sub classes, CEMP-$s$ stars with their strong enhancements in $s$-process elements 
 are mainly  found to be members of a binary system \citep{aoki2007apj,Starkenburg2014,hansentt2016aa2}. This favors 
 the idea that their carbon-enhancement is due to mass transfer from a  companion,  which was once in its Asymptotic Giant Branch (AGB) phase and has since faded into a white dwarf. Thus, the carbon- and $s$-process-rich material had been accreted onto the surface of the presently observed star,
 which  does not necessarily reflect the properties of the genuine ISM from which it was born. 

CEMP-no stars, on the other hand, are often found not to be
associated with binaries  (according to \citet{hansentt2016aa}, 
4 out of 24 stars in the radial velocity monitoring show binarity) and, as such, their abundance patterns are those imprinted by the ISM. 
\citet{hansentt2016aa} stated that CEMP-no stars are {\em bona fide} 
second generation stars, bearing the primordial remnants of enrichment from an early generation of stars. 
The frequency of CEMP-no stars increases with decreasing metallicity, and below an  [Fe/H] of approximately 
$-3.5$ dex almost all C-rich stars  are of the CEMP-no class
\citep{bonifacio2015aa,cjhansen2016}. 
The abundance pattern observed in  CEMP-no stars matches well with  
models of primordial faint Supernovae (SNe) that experienced mixing and fall-back, 
 and models of zero-metallicity-spin stars having high rotational velocity \citep[e.g.,][]{Umeda2003,Meynet2006,Kobayashi2011,Ishigaki2014}. 
In this regard, \citet{Yoon2016} and \citet{cjhansen2016} suggest that probably 
more than one class of first-generation  progenitors is required to account for the abundance patterns of CEMP-no stars.

Recent studies have uncovered several carbon-rich stars in dwarf spheroidals (dSph) and ultrafaint dwarf satellites around the MW 
\citep{frebel2010apj,lai2011apj,norris2010apj2,tuc-ji2016apj}.
Broadly consistent with the properties of CEMP stars in the Galactic halo, these objects were found across the entire metallicity range down to [Fe/H]$\sim$$-3.5$ dex
and  as metal rich as $-2.0$ dex and above, with varying C-enhancements from moderate values of [C/Fe]$\sim$0.5 up to 2.3 dex,  
and drawn from the CEMP-no and CEMP-$s$ subclasses. 
Moreover, these stars have now been found and analyzed in detail 
in the more luminous dSphs such as Sculptor and Sextans \citep{honda2011pasj,skuladottir2015aa,salgado2016mnras}. Here, the latter works 
stand out in that they  provide the first carbon-rich 
stars within the metal-rich tails of the galaxies' metallicity distributions (at [Fe/H]=$-2$ and $-1$ dex). 
Even though all of the more luminous dSphs contain old stellar populations \citep{Grebel1997}, none of them appear to show any CEMP population in the 
extremely metal-poor regime below $-$2 dex, owing to the 
shift in the metallicity distribution function towards the metal-rich side with  increasing galaxy luminosity \citep{Kirby2011,salvadori2015mnras}.  

Amongst the classical dSphs,  Carina is of  special  interest because of its unusual, episodic star formation history, where each of the stellar populations 
has also experienced distinct chemical enrichment \citep{Mould1983,Smecker-Hane1994,Monelli2003,Tolstoy2003,koch2008aj}.
\citet[][herafter ALW]{Azzopardi1986} detected ten carbon-rich (CH- and C-) stars in Carina.
Two of those (ALW-6 and ALW-7) were analyzed by \citet{abia2008aa}, with a particular focus on the  origin of the $s$-process elements
in these stars. At [Fe/H]=$-1.8$ to $-2$ dex and [Ba/Fe]$>$1.9 dex, both qualify as CEMP-$s$ stars, where mass transfer 
of processed material acted as the source of the abundance enhancements. 
Conversely, no CEMP-no stars have been identified in Carina to date.

In this paper,  we report  on the abundance analysis of the first CEMP-no star in  Carina (ALW-8), 
which shows no enhancement in neutron capture elements. Another member of the ALW-sample  (ALW-1) 
could be shown to be a CEMP-$s$ star; however its binarity  inhibited a detailed abundance analysis. 
This paper is organized as follows: In Section~2, we briefly recapitulate the observations and data for the stars, and 
in Sect.~3 we describe in detail our abundance analysis for ALW-8, the results of which are presented in Sect.~4. 
In Sect.~5, the resulting abundances are discussed in terms of the origin of the elements in this star, before we conclude our 
measurements in Sect.~6. 
\section{Observations and data reduction}
The data were obtained in the course of the ESO Large Programme
171.B-0520 (PI: G.~F.\ Gilmore) that aimed
at studying the kinematic and chemical characteristics of Local Group
dSphs \citep{koch2006aj, koch2008aj}. 
 Along with high-resolution ($R$$\sim$40000) spectra of ten red giants that
were observed with the Ultraviolet and Visual Echelle Spectrograph (UVES)
at the ESO/VLT in multi-object mode, 
two of the targets turned out to be carbon-rich stars as was obvious from
their strong molecular bands.
Coincidentally, these were part of the
ALW sample, namely ALW-1 and ALW-8.
The C-rich nature of ALW-8 was also noted by \citet{venn2012apj}, but discarded from their further analysis.
Our observations and data reduction for these two stars
are identical to those of the red giant sample of \citet{koch2008aj}, to
which we refer the reader
for details.  The resulting UVES spectra cover a full wavelength range from 4800 to 6800\AA.

Fig.~1 shows the location of the stars in a color magnitude diagram with
photometry taken from the
ESO Imaging Survey \citep{nonino1999aas}, and their general properties are  listed in Table~1. 
\begin{table}[b]
\caption{Photometric, kinematic, and atmospheric parameters of the target stars}
\centering          
\begin{tabular}{cccc}
\hline\hline       
Property & ALW-1 & ALW-8 & Reference\tablefootmark{a} \\
\hline
$\alpha$ (J2000.0) & 06:41:08.58 & 06:41:46.27 & 1 \\
$\delta$ (J2000.0) & \llap{$-$}50:47:50.1\phantom{0} & \llap{$-$}50:58:55.9\phantom{0}  & 1 \\
V & 17.68 & 17.88 & 2 \\
I  & 16.33 & 16.41 & 2 \\
J & 15.235 & 15.465 & 3 \\
H & 14.557 & 14.778 & 3 \\
K & 14.295 &  14.509 & 3\\
E(B$-$V) & 0.053 & 0.064 & 4\\
v$_{\rm HC}$ [km\,s$^{-1}$] & 226.2 / 256.2\tablefootmark{b}  & 223.3 & 5  \\
T$_{\rm eff}$ [K] & \ldots & 4150  & 5 \\
log\,$g$ & \ldots & 1.00   & 5 \\
$\xi$ [km\,s$^{-1}$] & \ldots & 2.3   & 5 \\
$[$Fe/H$]$ & \llap{$\sim -$}2.8 & \llap{$-$}2.5  & 5 \\
C/O & \ldots & 1.5 &   5\\
$^{12}$C/$^{13}$C & \ldots & 9  & 5\\
\hline
\hline
 \end{tabular}
\tablefoot{
\tablefoottext{a}{References: $1$: ALW; $2$: \citet{Walker2009}; 
$3$: 2MASS \citep{Cutri2003}; 
$4$: \citet{Schlegel1998}; 
$5$: This work.}
\tablefoottext{b}{Radial velocity from the two epochs of our observations.}
}
\end{table}
\begin{figure}
\centering
 \includegraphics[width=0.7\hsize]{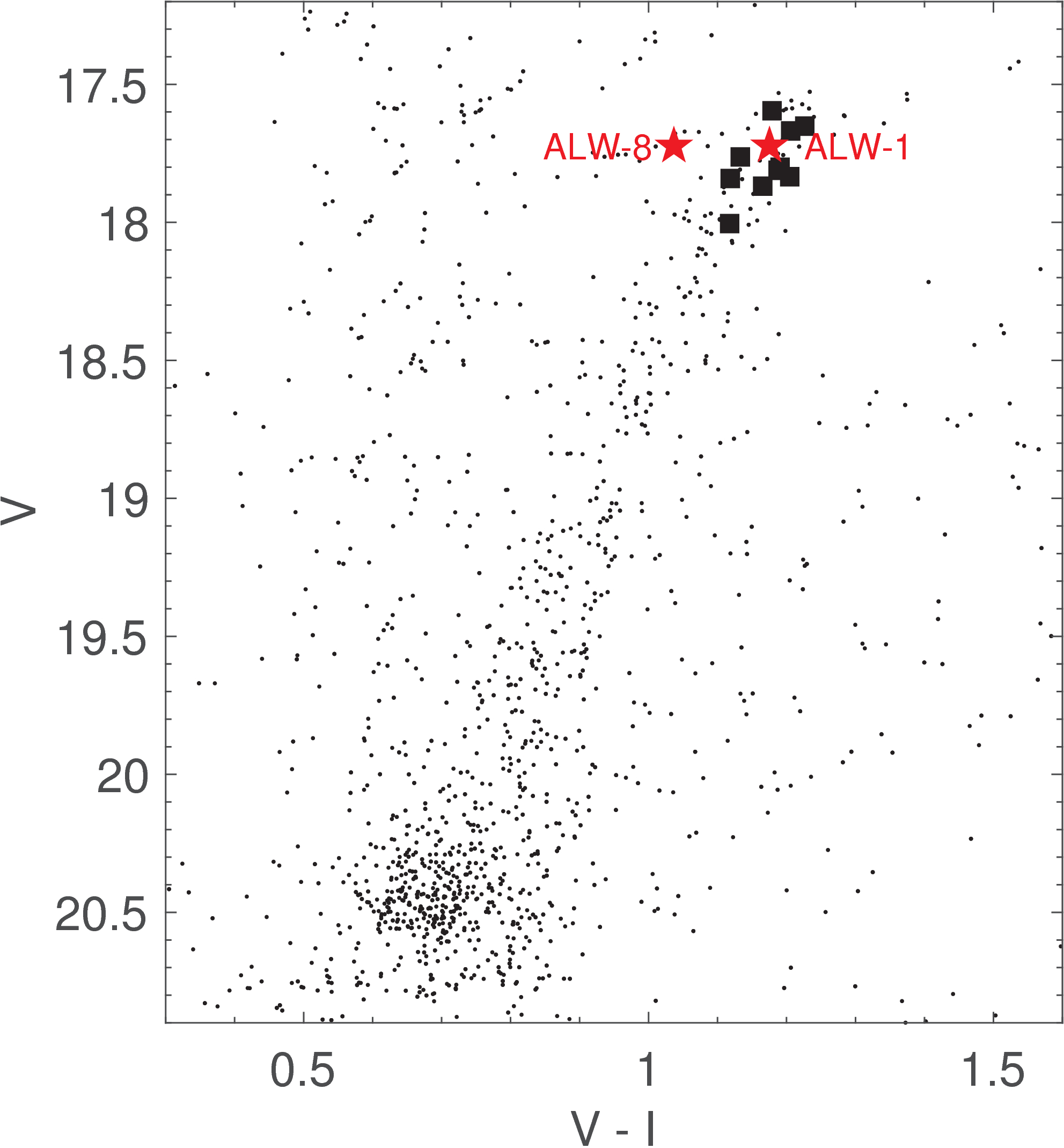}
\caption{Upper color magnitude diagram of Carina's central 10$\arcsec$, using photometry from the ESO Imaging Survey. 
Stars shown as black squares are the red giant targets
of Koch et al. (2008), while the filled red star symbols indicate the carbon stars from the present work. }
\end{figure}
\subsection{Radial velocity and binarity}
In order to correct for the shift in spectral lines, the  spectra of both stars were cross-correlated against a synthetic spectrum generated on the initial 
assumptions of temperature, surface gravity,  and the mean metallicity of Carina \citep{koch2006aj}.  
The resulting heliocentric velocity of ALW-8  of 223.3$\pm$1.0 km\,s$^{-1}$ is not only in excellent 
agreement with Carina's systemic velocity, but also agree well with the values of these stars found by \citet{Walker2009}. 

For  ALW-1, however, the two epochs of our observations returned velocities  deviating  by 30 km\,s$^{-1}$. Likewise,  the data of 
\citet[][and 2017, private communication]{Walker2009} and T.T. Hansen (2017, private communication) indicate significant radial velocity variations 
that suggest that ALW-1 is in fact part of a binary system.
Therefore,  we discarded the spectrum from further consideration due to the large uncertainties imposed by its low signal-to-noise ratio and difficulties in continuum placement due to 
veiling \citep[e.g.,][]{thompson2008apj}. 
A cursory analysis indicated a metallicity of $-2.8$ dex, a large [C/Fe] ratio of $\sim$1.4 dex, and a strong enhancement in $s$-process elements with a [Ba/Fe] ratio 
of approximately 2 dex. This shows that ALW-1 is likely to be a CEMP-$s$ star. 
\section{Abundance analysis}
Throughout our work, we used the spectral synthesis code TURBOSPECTRUM developed by \citet{Plez2012ascl}. 
To this end, we interpolated the final stellar atmospheric model from a
grid of model photospheres \citep{meszaros2012aj} in which the 
ATLAS9 and MARCS codes were modified with an updated H$_2$O linelist and with
  a wide range of carbon- and $\alpha$-element enhancements.
Local thermodynamic equilibrium (LTE) has been assumed for all species.

We adopted the Solar abundances from \citet{asplund2009araa} and Solar isotopic ratios were used for all the elements unless noted otherwise (see Sect.~4.3.). 
Our line lists  for atomic lines were assembled from the VALD database \citet{kupka1999aas} and details are given in Table~2. 
Hyperfine structure (HFS) has been accounted for for Li, Sc, Ba, and Eu, although the corrections
were negligible in the latter two cases. 
Finally, for the relevant molecules we employed the CH line list compiled by T. Masseron (private communication) and CN data from \citet{plez-cohen2005aa}.
\subsection{Equivalent-width measurements}
The  spectrum of the CEMP star ALW-8  shows a wealth of molecular features that contaminate the atomic lines. 
This severely hindered the measurements of precise equivalent widths (EWs)  for a large number of lines.
To this end, we constrained our analysis to such few lines that are devoid of contributions from molecular bands. 
In order to identify such unblended lines, we synthesized the full spectral range 
using the representative carbon, nitrogen, and oxygen abundances, and checked for those lines that remained free from molecular features.
In practice, EWs were measured by fitting Gaussian profiles to the features using  IRAF's {\em splot} task.
The uncertainties in the measurements were determined via the revised Cayrel formula \citep{cayrel1988iaus, battaglia2008mnras}.
The final linelist is given in Table ~A.1 in the appendix, which is also available in electronic form via the CDS.
\subsection{Stellar parameters}
The effective temperature of ALW-8 has been estimated using various color indices from the literature (Table~1), and using the color-transformation of 
Alonso et al. (1999). To this end, we adopted an initial metallicity of $-2.5$ dex and 
a reddening of E(B$-$V) = 0.064 \citep{Schlegel1998}. 
At T$_{\rm eff}$=(4150$\pm$40 K, 4200$\pm$170 K, 4240$\pm$125 K) from the V$-$K, J$-$H, and J$-$K colors, respectively, all photometric 
values are in very good agreement within the errors. In the following, we adopt the value from the V$-$K color of  4150$\pm$40 K as the stellar temperature.
Given the broad color range spanned by this index, this is the most reliable indicator, and it also corresponds to the error-weighted mean of all values above. 
This also yielded a flat trend of abundance of Fe I lines with excitation potential. 
By demanding that the slope of the 
iron abundances from the neutral lines with excitation potential do not change by more than one standard deviation of the Fe\,{\sc i} line abundances, 
we placed an error of 100 K on the spectroscopic temperature.

The surface gravity, log\,$g$ was derived by fitting a grid of synthetic spectra to the  Mg triplet lines around 5170 \AA, where we adopted 
the temperature derived above as the model's T$_{\rm eff}$. The best-fit model thus found is indicated in Fig.~2. 
Acceptable fits, in particular accounting for the Mg triplet's wings, 
were still obtained within variations of $\pm$0.25 dex, which we adopted as the error on our log\,$g$ determination.
\begin{figure}
\centering
 \includegraphics[width=1\hsize]{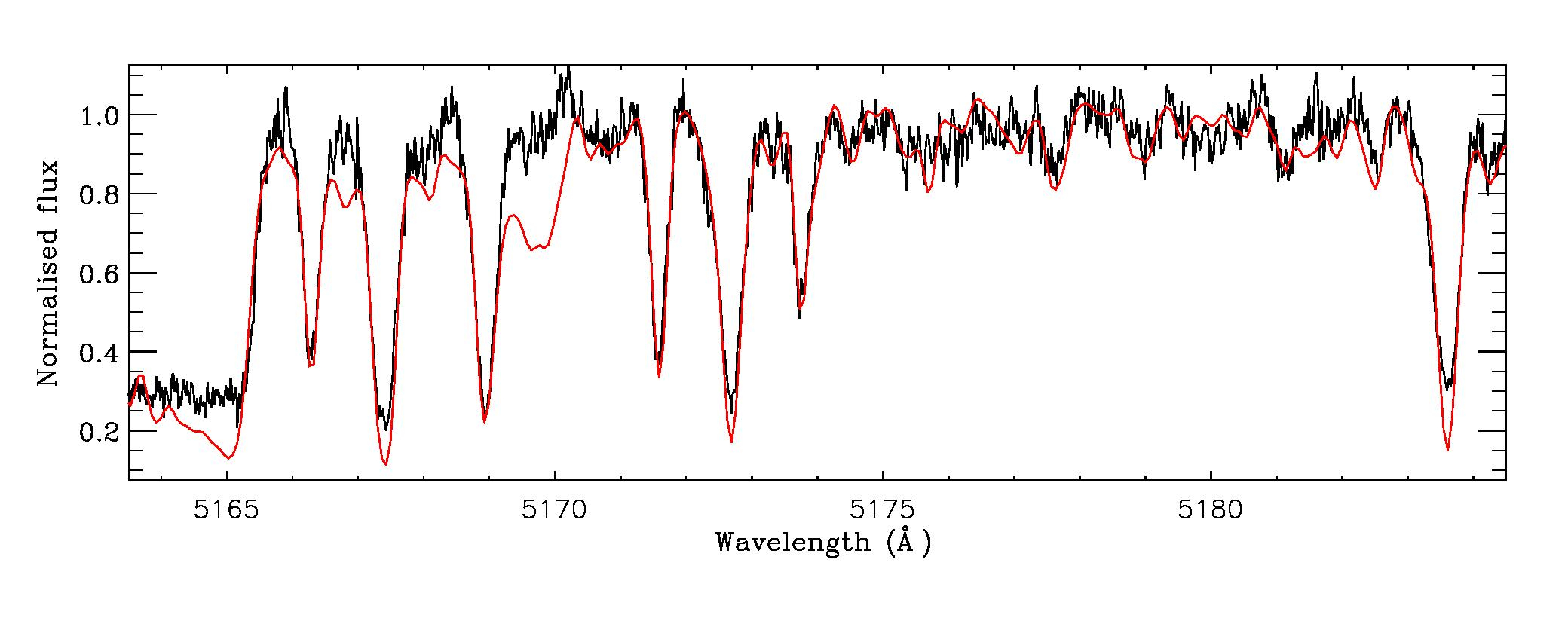}
 \includegraphics[width=1\hsize]{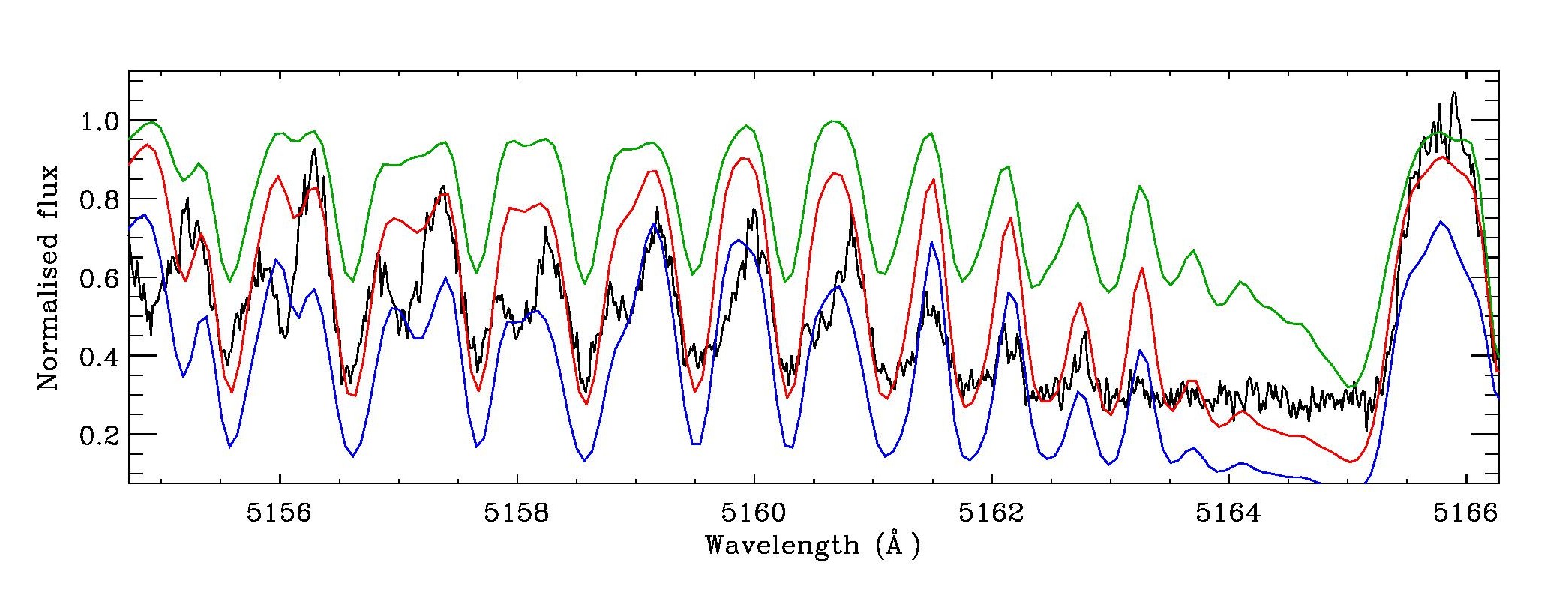}
\caption{Top panel: Region of the Mg I triplet in ALW-8 (black line). The best-fit synthetic spectrum is plotted in red. 
The additional feature at 5170\AA\ is due to the chosen carbon abundance. 
Bottom panel: Synthesis of the $C_{2}$-band in ALW-8. The best-fit spectrum is shown in red ([C/Fe]=+1.39), while the blue and green lines have been computed for 
carbon abundances that differ by $\pm$0.2 dex.} 
\end{figure}
Finally, the microturbulent velocity, $\xi$, was set by removing any trend of  the abundance from the Fe\,{\sc i} lines 
with the reduced width, $\log$(EW/$\lambda$). The resulting  value was found to be $\xi$= 2.3 km\,s$^{-1}$ with 
an error of 0.2 km\,s$^{-1}$, as determined by the point where the Fe\,{\sc i} abundance did not change by more than 1$\sigma$.
While fixing the surface gravity, the C-abundance  was iteratively changed from log $\epsilon$(C)= 6.42  to log $\epsilon$(C) = 8.02 in steps of 0.2 dex to converge on 
 the best-fit parameters.
 All stellar parameters obtained in this way are listed in Table~1. 
\section{Abundance results}
All abundance results are listed in Table~2, together with our adopted Solar values    from   \citet{asplund2009araa}. 
In Fig.~3, our results are compared to literature measurements in the Galactic disks \citep{Koch2002,Bensby2014,Battistini2016} 
and halo  \citep{Roederer2014}, where we also included data for  C-normal stars in Carina \citep{koch2008aj,venn2012apj,fabrizio2015aa}, 
other C-rich stars in MW dSph satellites, including Carina's ALW-6 and -7 \citep{abia2008aa}, and the sparse available data from a few ultrafaint galaxies
\citep{Geisler2005,frebel2010apj,norris2010apj,honda2011pasj,skuladottir2015aa,salgado2016mnras,tuc-ji2016apj,ret-ji2016apj}. 
We note that the Galactic data have not been preselected to show exclusively CEMP stars, but they are, rather, representative of the 
metal-poor MW field population. Similarly, for the faint dSphs, we only show those 
stars that have several key elements derived from high-resolution spectra, while the overall detection rate of carbon excesses from 
low-resolution spectra without further chemical follow-up is larger \citep[e.g.,][]{lai2011apj}. Moreover, a few of the 
 CEMP-stars in the faintest dSphs often fall within  the CEMP category by a margin of their [C/Fe] ratio \citep{aoki2007apj}.
\begin{table}[htb]
\centering
\caption{Abundance results.}
\begin{tabular}{ccccccc}
\hline\hline
{Species} & {log $\epsilon_{\odot}$} & log $\epsilon$   & [X/Fe] & $\sigma_{log \epsilon}$ & N & $\sigma_{sys}^{tot}$\\
\hline
Li\,{\sc i}       & 1.05 &               0.00 &              1.45 &   0.30 &  syn & 0.32 \\
C (C$_{2}$)       & 8.43 &               7.32 &              1.39 &   0.20 &  syn & 0.15 \\
N (CN)            & 7.83 &               6.92 &              1.59 &   0.30 &  syn & 0.14 \\
O\,{\sc i}        & 8.69 &               7.08 &              0.89 &   0.30 &  syn & 0.10 \\
Na\,{\sc i}       & 6.24 &  4.62$_{\rm NLTE}$ & 0.88$_{\rm NLTE}$ &   0.12 &         2 & 0.09 \\
Mg\,{\sc i}       & 7.60 &               5.40 &              0.30 &   \ldots &    1 & 0.13 \\
Ca\,{\sc i}       & 6.34 &               4.26 &              0.42 &   0.27 &         5 & 0.13 \\
Sc\,{\sc ii}      & 3.15 &               1.17 &              0.52 &   0.09 &         3 & 0.16 \\
Ti\,{\sc i}       & 4.95 &               2.75 &              0.30 &   0.37 &         7 & 0.18 \\
Ti\,{\sc ii}      & 4.95 &               2.85 &              0.40 &   0.15 &         3 & 0.16 \\ 
Cr\,{\sc i}       & 5.64 &               2.94 &    \llap{$-$}0.20 &   0.29 &         5 & 0.16 \\
Mn\,{\sc i}       & 5.43 &               2.63 &    \llap{$-$}0.30 &   0.10 &         2 & 0.19 \\
Fe\,{\sc i}       & 7.50 &               5.00 &              0.00 &   0.12 &        11 & 0.12 \\
Fe\,{\sc ii}      & 7.50 &               4.99 &              0.01 &   0.10 &      2 & 0.19 \\
Ni\,{\sc i}       & 6.22 &               3.56 &    \llap{$-$}0.16 &   0.15 &         3 & 0.11 \\
Zn\,{\sc i}       & 4.56 &               2.24 &              0.18 & \ldots &      1 & 0.12 \\
Y\,{\sc ii}       & 2.21 &               0.00 &              0.29 & \ldots &  syn & 0.11 \\ 
Zr\,{\sc ii}      & 2.58 &     \llap{$<$}0.50 &    \llap{$<$}0.42 & \ldots &  syn & 0.42 \\
Ba\,{\sc ii}      & 2.18 &     \llap{$-$}0.86 &    \llap{$-$}0.55 & \ldots &  syn & 0.19 \\
Eu\,{\sc i}       & 0.52 & \llap{$<$ $-$}1.50 &    \llap{$<$}0.48 & \ldots &  syn & 0.86 \\
\hline\hline                      
\end{tabular}
\label{tab:abund}
\end{table}
\begin{figure*}[!htb]
\centering
\includegraphics[width=1\hsize]{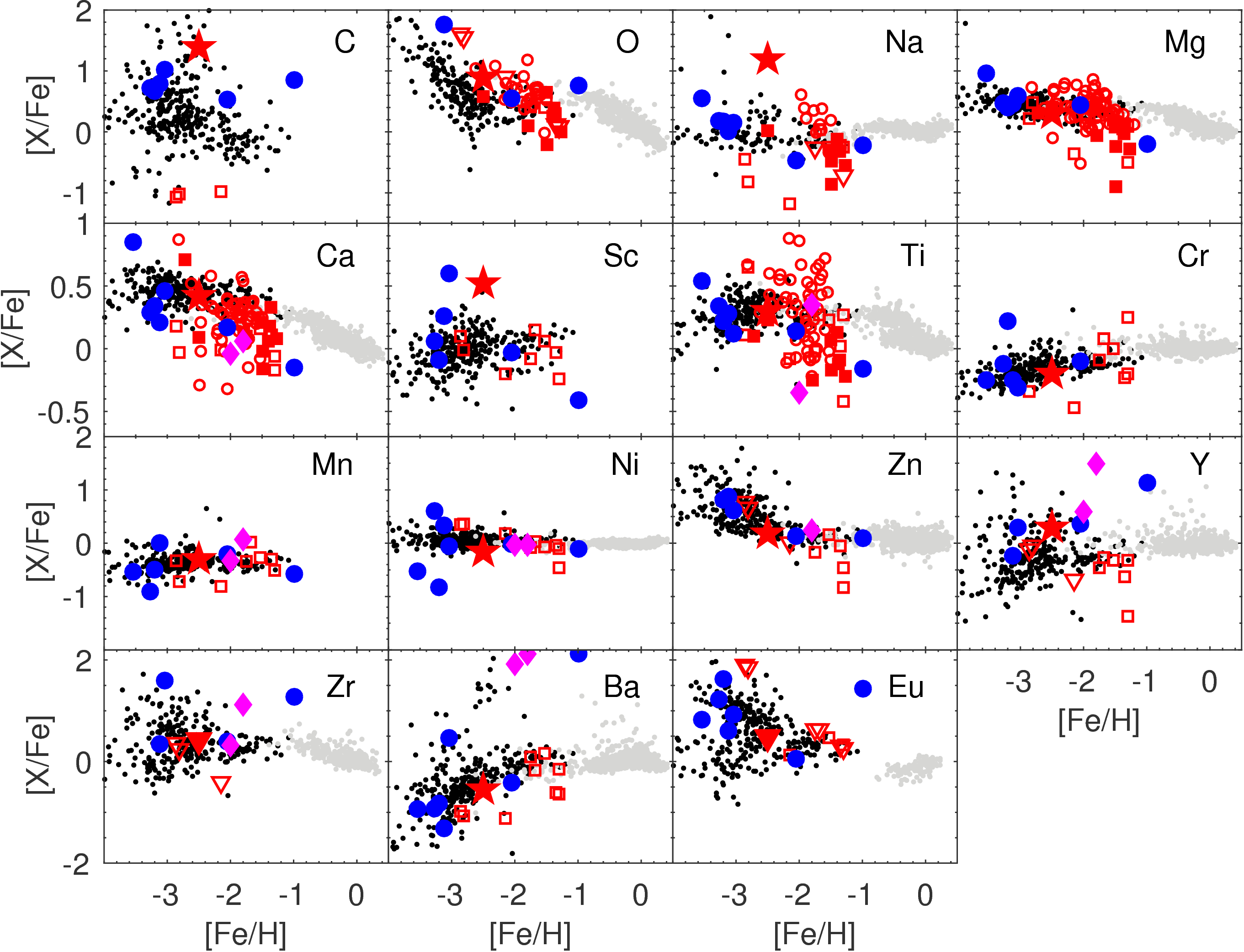}
\caption{Abundance ratios for ALW-8 (red star symbol  and, for the case of upper limits, filled red triangles) in comparison with MW and dSph samples from the literature. Data 
are from \citet{Roederer2014} for halo stars (black dots);  \citet{Bensby2014}  for disk stars (gray points), except for Zr \citep{Battistini2016} and Eu \citep{Koch2002}. 
Carina field star abundances were taken from \citet[][filled red squares]{koch2008aj}; \citet[][open red squares]{venn2012apj}; and \citet[][open red circles]{fabrizio2015aa}, while  open red triangles
 indicate upper limits from these studies.
Blue points designate C-rich stars in luminous and ultrafaint dSphs \citep{Geisler2005,frebel2010apj,norris2010apj,honda2011pasj,skuladottir2015aa,salgado2016mnras,tuc-ji2016apj,ret-ji2016apj}. Finally, the CEMP-$s$ stars in Carina from \citet{abia2008aa} are shown as magenta diamonds.
}
\end{figure*}
\subsection{Abundance errors}
In order to quantify the statistical error due to contributions from uncertainties in EWs and atomic parameters, we
list in Table~2 the 1$\sigma$ line-to-line scatter, $\sigma_{\log\epsilon}$ and the number of lines on which each element's abundance was based.
For those few cases, where the abundance could only be derived from less than three lines, this random error was adopted as  
0.10 as an empirical, conservative upper limit. 
 For the abundances derived via spectrum synthesis, the quoted abundance uncertainty was determined via the goodness of the least-squares fit. 

Systematic errors in our abundances were derived by computing new models, where each stellar parameter (T$_{\rm eff}$, log\,$g$, $\xi$) was
varied by a fixed amount ($\pm$100 K, $\pm$0.25 dex, $\pm$0.2 km\,s$^{-1}$), and from those, new abundances were computed in an identical manner as before. 
The resulting line-to-line change upon these parameter variations is listed in Table~A.1, and the combined effect  was determined by 
adding the contributions in quadrature. These systematic uncertainties are listed in the last column of Table~2.
\subsection{Iron}
Many  iron lines are severely blended with  CN and C$_{2}$ molecular features, but we were able to 
measure the EWs of 11 Fe\,{\sc i} lines that are free of blends.
In addition, we synthesized the full spectral range in its entirety in order to assess the accuracy of the abundance analysis and the influence of such blending. 
As a result, the values from the EW analysis and spectral synthesis are in excellent agreement.  We thus report the Fe-abundance of ALW-8 as
[Fe/H] = $-2.50\pm 0.04$.

The covered spectral range also allowed us to determine the abundance from two unblended Fe\,{\sc ii} lines.
The resulting abundance from these lines indicates an excellent ionization equilibrium in that [Fe\,{\sc i}/{\sc ii}]=0.01$\pm$0.13, 
which also renders our surface gravities from the  synthetic grid fitting reliable.  
\subsection{Carbon, Nitrogen, Oxygen, and  $^{12}$C/$^{13}$C ratio}
The carbon abundance of this star has been derived by fitting the C$_2$ molecular band heads at 5164 \AA\ and 5635 \AA\ (Fig.~2).
Both  features yielded the same abundance of  log $\epsilon(C)$ = 7.3$\pm$0.2 dex.  
The isotopic ratio $^{12}$C/$^{13}$C was derived by fitting the $^{13}$C contribution to the band at 5634 \AA, resulting in a ratio of 9,
 which is consistent with  other CEMP-no stars
\citep{sivarani2006aa,aoki2007apj}. The equilibrium ratio for CNO-cycled material is $\sim$4 and the  
 value in ALW-8, resulting from the production of $^{13}$C in the CN cycle, indicates a high level of processing, in line with 
 the evolved nature of this star  \citep{charbonnel1998aa,gratton2000aa}. 
Since the strong CN band at 4215 \AA\ is not covered by our spectrum,  
we  used  CN  lines 
 in the wavelength range from 5635--6700\AA~by iteratively changing the nitrogen abundance of the 
synthetic spectra by 0.2 dex and fitting this entire spectral region. This yielded a best-fit value of 
[N/Fe]=1.6.

Finally, the oxygen abundance of ALW-8 was derived   by using the   [O I] lines at 6300, 6363 \AA~(Fig.~4). 
Since these lines are blended with CN  features, a slight change in the nitrogen abundance affects the derived oxygen abundance. 
Thus,  the carbon abundance was first fixed to be the value obtained above from the $C_{2}$ band, after which  the oxygen and nitrogen abundances
were  iteratively adjusted to fit the [O I] lines, without affecting  other CN features. 
The resulting O-abundance from both lines differs by 0.3 dex owing to  the iterative process involving both N and O, and we adopt this difference as our final 
uncertainty on the reported nitrogen and oxygen abundances. 
\begin{figure}[htb]
\centering
\includegraphics[width=1\hsize]{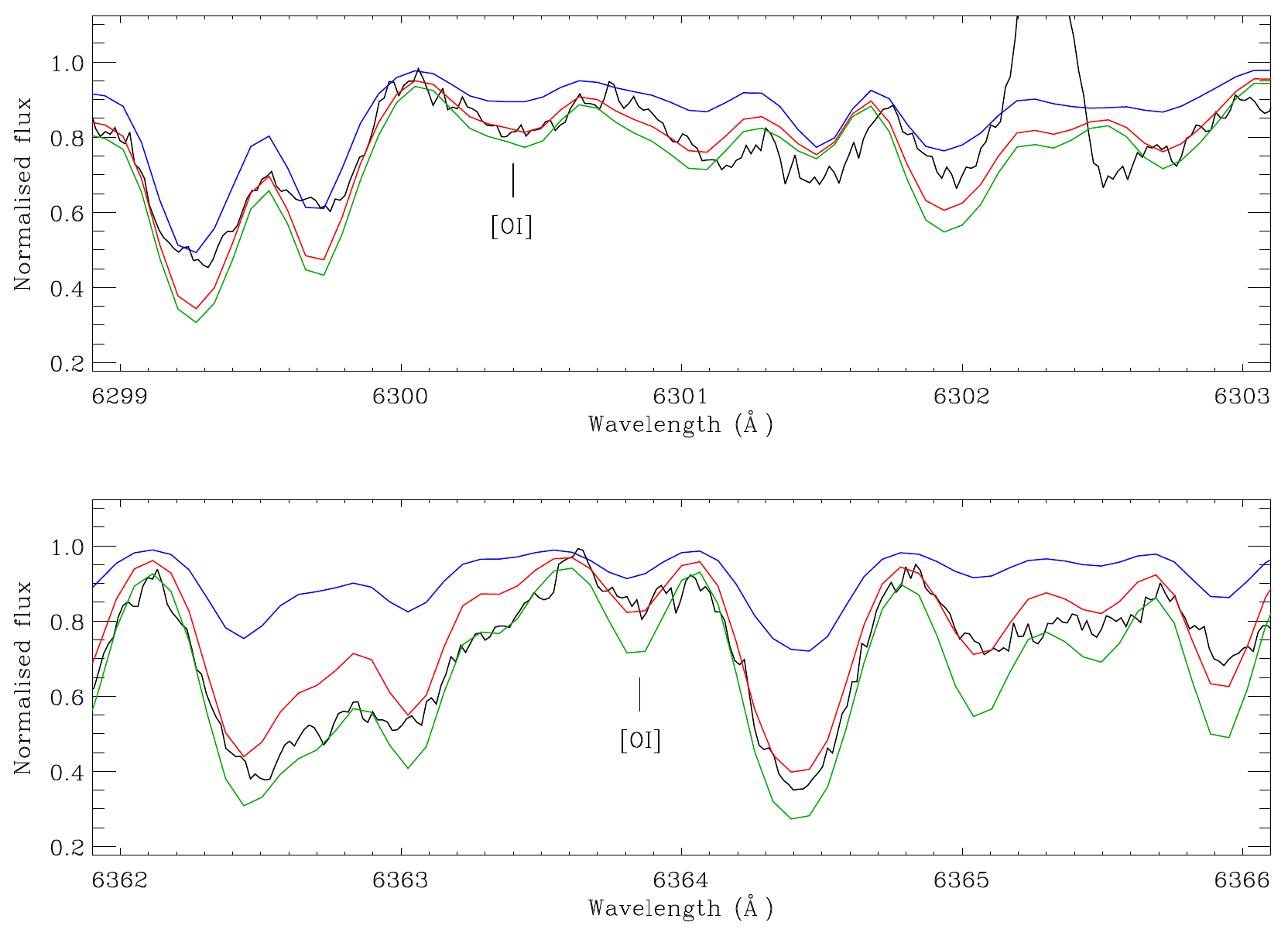}
\caption{Regions around the  [O I] lines in ALW-8. The best spectrum (red) was computed with  log $\epsilon$(O)=6.93 (top panel) 
and log $\epsilon$(O)=7.23 (bottom panel), respectively. 
The green and blue lines indicate syntheses corresponding to the range of uncertainty of $\pm$0.3 dex in oxygen abundance.
The poor fitting of CN features around [OI] comes solely from the 
change in the oxygen abundance.}
 \end{figure}
\subsection{Lithium}
The Li-abundance for the star was determined by synthesizing the resonance doublet at 6707 \AA\, which yielded A(Li)=0.0. 
Due to the presence of prominent CN bands in this region, affecting the continuum placement, we assigned a fitting uncertainty of 
0.3 dex to this result.
The low value for A(Li) we found is consistent with the star being an evolved giant, in which Li is easily destroyed at the high interior temperatures
that it is exposed to due to the convective mixing of material \citep[e.g.,][]{gratton2000aa,lind2009,skuladottir2015aa}. 
In contrast, unevolved CEMP stars show Li-abundances that are lower than the standard plateau value of 2.2 dex \citep{Spite1982,Sbordone2010,Masseron2012,bonifacio2015aa}.
\subsection{$\alpha$-elements: Mg, Ca, and Ti}
Clean lines of Mg, Ca, and Ti are available in the spectrum of ALW-8 and the abundances of the respective elements have been determined from their EWs. 
The  Mg I triplet lines at 5172 \AA\  and 5183 \AA\ are very strong (at EWs $>$300 m\AA) and saturated; conversely, 
the line at 5711\AA~ is strongly blended with  molecular features and thus unusable  for an abundance determination. 
Hence, we base our measurement of the Mg-abundance on the unblended
line at 5528 \AA\ at moderate strength (120 m\AA), resulting in a [Mg/Fe] ratio of 0.3 dex. 
 
 A poor mixing has been found  between Carina's old and intermediate-age populations in terms of their Mg abundances \citep{shetrone2003aj,koch2008aj, lemasle2012aa,venn2012apj}.  Here, older stars with systematically lower Fe abundances show depleted [Mg/Fe] ratios, 
 whereas the intermediate-age population  is more strongly  enhanced in Fe and the $\alpha$-element Mg. 
This left \citet{venn2012apj} to conclude that the second phase of star formation in Carina occurred out of gas 
 that was already pre-enriched in the $\alpha$-elements. Moreover, the gas out of which the older stars formed was inhomogeneously mixed itself, leading to a broad
 spread in the [Mg/Fe] ratios at low metallicities.
 At its low [Fe/H] and the elevated [Mg/Fe], ALW-8 is clearly part of Carina's old population.
   
Five Ca\,{\sc i} lines have been used to determine a halo-like  [Ca/Fe]  ratio of  0.42 dex, while, for Ti, 
eight and three neutral and ionized lines were clean and detectable, respectively. These indicate 
that ionization balance is matched very well within the uncertainties, at [Ti\,{\sc i}/{\sc ii}]=0.10$\pm$0.18 dex.
No Si lines are  present in the available range of the spectrum.

The straight average of the three $\alpha$-elements amounts to an enhancement of [$\alpha$/Fe] = 0.36 dex.
Even though the trends indicating the poor mixing between the components is less pronounced in the other $\alpha$-abundances, 
there is still considerable spread in Ca and in particular Ti \citep{fabrizio2015aa}. ALW-8 behaves rather halo-like in its enhancements.
While this overlap is also seen for several of the CEMP stars in other dSphs, two of these stand out clearly. One is the most metal-poor of these objects, 
\citep[the CEMP-no star in Segue~1;][]{norris2010apj}, which is $\alpha$-enhanced to almost one dex.
Secondly, the most metal-rich C-rich star in Sculptor, at $-1$ dex
\citep{Geisler2005}, shows subsolar [$\alpha$/Fe] ratios; although we note that it  represents the typical [$\alpha$/Fe] ratio of stars at similar metallicity in Sculptor. 
Also the CEMP-$s$ stars ALW-6 and -7 are towards the low-$\alpha$ tail 
of Carina's distribution. 
The fact that ALW-8 and the majority of the other C-stars follow the overall trends in the $\alpha$-elements 
very well  is not surprising, since the nucleosynthesis of the $\alpha$-elements is
decoupled from the channels responsible for the C-overabundances.  
\subsection{Odd-Z elements: Na and Sc}
Since the Na D lines are too strong (EW $\sim$ 300 m\AA\ )  for a meaningful abundance analysis, we
relied on the  weaker absorption lines at 5682 and 5688 \AA. 
We have incorporated NLTE corrections from \citet{Lind2011} 
and the resulting NLTE [Na/Fe] ratio is high, at 0.9 dex, skimming the upper distribution of halo stars and Carina stars.  
\citet{fabrizio2015aa} found that a correlation between Na and O in their Carina stars matched well with that in MW  halo field stars, 
 supporting the  similarity between the chemical enrichment history of the  MW halo and Carina stars \citep{idiart2000apj,Geisler2007}. 
Unfortunately, no Na abundances were derived for the old stars in their sample.  
While, at first glance, the correlation noted by \citet{fabrizio2015aa} appears to be driven by a single, low-Na, low-O star, 
the very high Na and O abundances in ALW-8 provide a unique match to this trend at the other extreme, bolstering the presence of
this correlation beyond the intermediate-age population.  This suggests that, apart from an O-enhancement in the CNO processing that this star experienced, 
the gas out of which it formed was already strongly pre-enriched in O and Na via previous Type-II SNe events. 

No  lines of Al and K  could be measured in the spectrum and all available V-lines are too heavily blended with molecular features 
to infer any abundance.

The Sc-abundance we measured from three generally strong and unblended lines yields a  
high [Sc/Fe] ratio of 0.52$\pm$0.05 dex, which is significantly larger than in halo-, Carina-, and CEMP-stars at similar metallicity. 
These reference samples typically fall within $\pm$0.2 dex of the Solar value. 
One exception is the CEMP-$s$ star S15-19 in the Sextans dSph \citep{honda2011pasj}, which, however, shows neutron-capture 
element patterns that are fully consistent with the AGB mass-transfer scenario, and no explanation of the high Sc/Fe ratio was offered.
Moreover, the horizontal branch star CS 29497-030 in the sample of \citet{Roederer2014}, at the same [Fe/H], shares almost identical abundances of all  the light elements in common, while  its neutron-capture elements are highly enhanced, classifying that object as a CEMP-$r/s$ star.
A high Sc ratio could be indicative of a high-energy SN event and/or a high electron fraction in the nucleosynthetic environment \citep{Tominaga2007}. 
However, there is little evidence for this acting in the neutron-capture element patterns seen in ALW-8 (Sect~4.8).
Furthermore, the NLTE corrections for Sc in warmer stars are still largely unknown, hampering a detailed model comparison to pin down the origin of
the Sc overenhancement  \citep{CJHansen2011}.
\subsection{Fe-peak elements: Cr, Mn, Ni}
As the Fe-peak elements are co-produced with iron and their nucleosynthesis is decoupled from that in charge of the carbon-enrichment and heavy element patterns
in the CEMP stars, the [Cr/Fe], [Mn/Fe], and [Ni/Fe] ratios bear little surprise. ALW-8 is fully compatible with the (low-scatter) trends in the MW halo and 
those seen in other CEMP stars of either class, as well as with the abundances found in Carina stars, albeit their larger scatter. 
We did not attempt to correct our Cr and Mn abundances for NLTE effects, but we note that these would have no impact on the normality of ALW-8 in this abundance space. 
\subsection{Neutron-capture elements: Zn, Y, Zr, Ba, Eu}
\begin{figure}[htb]
\centering
\includegraphics[width=1\hsize]{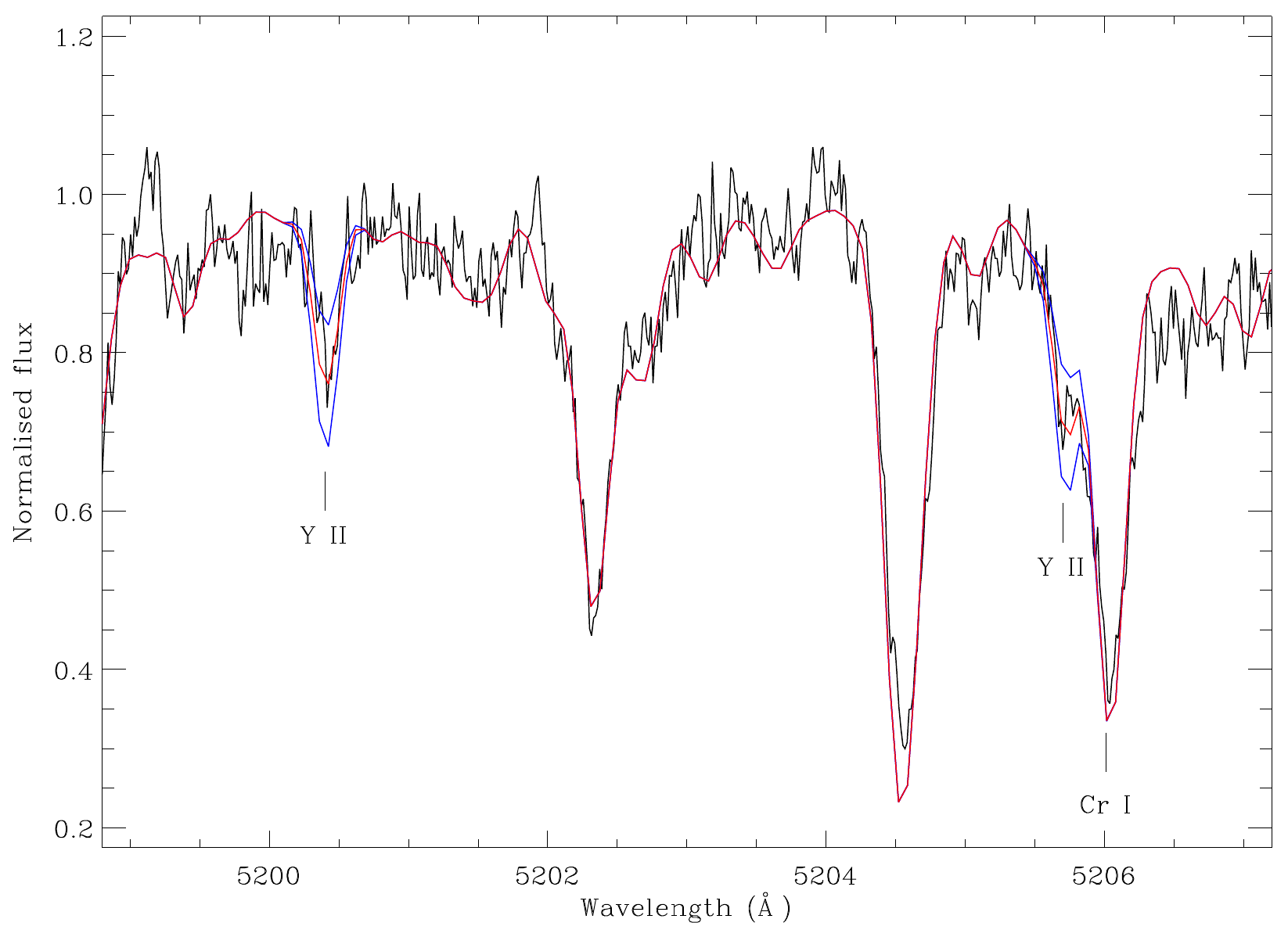}
\caption{Spectrum of ALW-8 around the Y\,{\sc ii} lines at $\lambda\lambda$5200, 5206\AA~(black line), illustrating our spectral fitting. 
The best synthetic spectrum using [Y/Fe] = 0.26 is plotted as a red line. 
The spectra shown in blue  differ by $\pm$ 0.3 from the best-fit value.}
\label{fig:y_57}
\end{figure}

The [Zn/Fe] ratio in ALW-8 and other CEMP stars is compatible with those in metal-poor halo stars, showing little scatter
in the samples. Furthermore, there is no systematic difference to be seen between  CEMP-no and -$s$ stars, which is to be expected as
Zn is not significantly affected by $s$-process nucleosynthesis \citep{Timmes1995,Kobayashi2006ApJ}. 
However, C-normal stars in Carina show a large scatter towards intermediate metallicities, where their 
low-Zn and Fe-peak abundances indicate a lack of hypernovae, that is,  energetic SNe that would lead to an overproduction of those elements \citep[see also][]{Hanke2017}. 
Amongst the first-peak $n$-capture elements, only Y and Zr abundances could be determined. 
For Y, we employed the lines at 5200 and 5206 \AA~(Fig.~\ref{fig:y_57}). While the latter feature is blended with a Cr I line, it can still be used for an abundance estimate, though with 
a  larger error bar. As a result, we find an elevated [Y/Fe] ratio of 0.3 dex, which still overlaps with halo stars owing to the large scatter found in that MW component.
We will further investigate this element in Sect.~5.2.
Unblended Zr lines could not be  detected in the spectrum and  we placed an upper limit on the Zr abundance by synthesizing the  lines at 6127 
 and 6134 \AA, which are   blended with CN bands; this   yields an  upper limit of [Zr/Fe] $<$0.42. 

Barium is the only heavy neutron-capture element with several strong lines 
present in the spectrum. Four Ba II lines were detected in the spectrum of ALW-8  at 4934.1, 5853.7, 6141.7, and 6496.9 \AA. 
However, the reddest of these lines is  too blended with molecular feature to be considered in our analysis. 
The  remaining three  lines yield a depleted value of [Ba/Fe]=$-0.55\pm$0.20 that is in line with other metal-poor stars in Carina and the Galactic halo. 
The [Ba/Fe] ratio shows a distinct bifurcation over a broad range in metallicity 
and both ALW stars in Carina studied by \citet{abia2008aa} fall on the high-Ba band that is made up of CEMP-$s$ stars 
in binaries. 

The weak Eu-line at 6645 \AA\  is  blended with CN molecular features 
and we were only able to place an 
upper limit of [Eu/Fe] $<$0.43 from spectral synthesis. 
\section{On the origin of elements}
Our abundance analysis of ALW-8 revealed that this CEMP star is enhanced in carbon as well as nitrogen. 
Considering its evolutionary phase, its $^{12}$C/$^{13}$C ratio, and the depleted Li abundance give a sense 
that this star has undergone first dredge up (FDU). 
Nitrogen is the result of a CN cycle acting during the  FDU. 
In this case, the original carbon in the star must have been even higher than what is measured now. 
Based on its low subsolar [Ba/Fe] ratio and considering the lack of any strong $r$-process enhancement, 
ALW-8  can  be unambiguously  classified as a CEMP-no star \citep{beers-christlieb2005ARA&A}. 
A mild enhancement is seen in the   light neutron capture element Y, which suggests the presence of a weak $r$-process activity.   
 Thus, in the following, we consider the origin of the enhancements seen in different element tracers in ALW-8. 
\subsection{Origin of carbon}
The strong C-enhancement of ALW-8 unambiguously confirms its nature as a CEMP star. 
 ALW-8 is in the RGB phase where the primordial carbon abundance has been reduced due to the CN cycling and the 
 presently observed abundance will be lower. 
While the moderate [C/N] ratio of $-$0.2 dex alone would  suggest that its surface composition has not been significantly altered by internal mixing during stellar evolution \citep{Spite2005,cjhansen2016}, other tracers of mixing paint a different picture. Amongst these are the strong depletion in lithium and 
the carbon isotopic ratio, where \citet{Spite2006A&A} suggest an upper limit of $^{12}$C/$^{13}$C $<10$ as a proxy for effective deep mixing.

To further assess the impact of such processes, we consulted the corrections of \citet{placco2014apj}, which account for 
the evolutionary status of C- and N-rich stars. As these model calculations indicate, the surface composition of stars like ALW-8 
has not been altered by more than $\sim$0.2 over the course of its evolution. Since, during the hot CNO cycles, N is strongly enhanced at the 
expense of lowering the carbon abundance, the high [C/Fe] we found in ALW-8 must have been a relic of external processes that enhanced the
primordial gas to the observed high levels.
 
Carbon in CEMP-no stars can have several origins (we refer to, e.g., \citealt{skuladottir2015aa} for a comprehensive overview). While 
the possibility of mass transfer from a AGB binary companion cannot be ruled out from a dynamical standpoint due to our single-epoch observations, 
 the lack of significant $s$-process material renders this unlikely; in fact, the majority of CEMP-no stars are not related to any binary mechanisms
 \citep{hansentt2016aa}.
 The high value of carbon in CEMP-no stars can be the result of primordial, low- or zero-metallicity faint SNe that underwent mixing and fall-back mechanisms \citep{Umeda2003}. Alternative enrichers can be fast rotating massive stars \citep{Meynet2006}. 
 However, both polluters produce excess abundances of N and O that are not produced to that extent in regular-energy SNe II. Indeed, our 
 nitrogen and oxygen measurements indicate a strong enrichment in those elements, which, however, is more likely to be a consequence of the CNO cycling and mixing
 that the star experienced. 
 Another piece of evidence against faint SNe being the sole producer of the high C is that they tend to 
 have very low yields of the odd-$Z$ Fe-peak elements such as Mn, which shows a regular, non-depleted value in ALW-8.

\begin{figure}[htb]
\includegraphics[width=1\hsize]{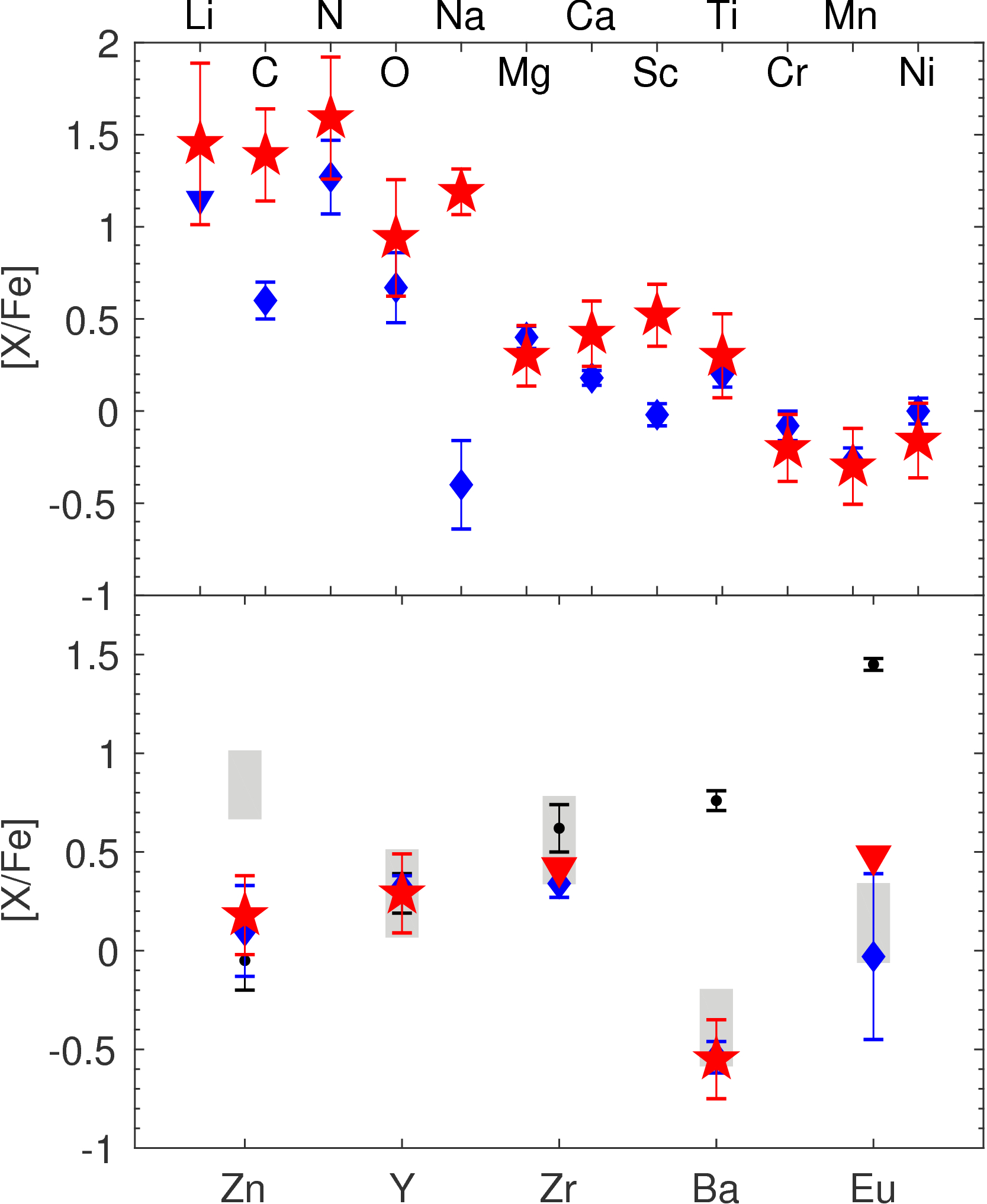}
\caption{Chemical element patterns for ALW-8  (red symbols) are shown for the lighter elements (top panel) and the neutron-capture element abundances (bottom panel).
For comparison, we overplotted the CEMP-no star in the Sculptor dSph \citep[][blue diamonds]{skuladottir2015aa}, and  the weak and main $r$-process stars HD 122563 \citep[][gray shades]{Honda2006} and CS 22892-052 \cite[][black dots]{Sneden2003}. 
 Here, all stars were placed on the same Solar scale. 
All ratios for the heavier elements were normalized to Y.}
\end{figure}
The very similar abundance pattern of the CEMP-no star in the Sculptor dSph (Fig.~6) prompted \citet{skuladottir2015aa} to 
propose that a mix of faint and regular SNe II is required to reproduce the high C without invoking any of the peculiarities present in the faint SNe models. 
The regular pattern observed in the $\alpha$-elements and the high Na abundance seen in ALW-8 agrees with such a scenario and we conclude that Carina has also experienced some degree of enrichment by faint SNe that was, however, diluted with yields from SNe II. 
Overall, we note the very good agreement of the [$\alpha$/Fe] and Fe-peak abundance ratios between the Sculptor star and our Carina CEMP star (top panel of Fig.~6).  
\subsection{Origin of neutron-capture elements}
At metallicities as low as ALW's [Fe/H] of $-2.5$ dex, the main source of the $n$-capture elements
is $r$-process nucleosynthesis without major contributions from the $s$-process in AGB stars \citep{Simmerer2004,sneden2008ARAA}.
Generally, the ratio of first-to-second peak $n$-capture elements, such as [Y/Ba], 
is systematically lower in dSphs than in MW halo, owing to their low star forming efficiency (Fig.~14 in \citealt{Tolstoy2009}). 
The opposite is seen in ALW-8, which exhibits a high [Y/Ba] of 0.84 dex, which is otherwise  seen in metal-poor 
halo stars below around $-$2.7 dex and also in the two most metal-poor Carina stars of \citet{venn2012apj} as well as in two other dSph CEMP stars. 
This reflects significant departures from the abundance patterns that are governed by the main $r$-process in that 
the first-peak elements (such as Sr, Y, Zr)  are systematically enhanced with respect to the heavier ones (Z$\ga$56). 
Such trends  prompted the need for an additional source of these elements via a weak $r$- or weak $s$-process \citep[e.g.,][]{Travaglio2004,Honda2006,Arcones2011}.
Also in ALW-8, only a low upper limit could be derived for Eu, which
serves as a tracer for main $r$-process.

Thus, in Fig.~6 (bottom panel), we place ALW-8 in the context of the $n$-capture processes by comparing the heavier element pattern to 
those in two metal-poor halo stars: HD 122563 
(Fe/H] = $-$2.82, [C/Fe]= -0.47)\citep{Spite2005,Honda2006}, 
which shows an archetypical weak $r$-process pattern, and
CS 22892-052 ([Fe/H]= $-$3.12, [C/Fe]= 0.89) \citep[][black dots]{Sneden2003} as a contender for enrichment via the main $r$-process. We note that all these reference stars have been 
normalized to the abundance of Y in ALW-8 in Fig.~6.
It becomes immediately clear that the low Ba and Eu abundances are incompatible with an overproduction in the standard $r$-process. 
 The scaled abundance ratio of Zn for HD 122563 deviates  from the other stars shown here, while we note that the absolute [Zn/Fe] ratios 
of these stars are consistent with one another. Thus, Zn appears comparably low; its origin  rather lies in complete Si-burning with possible contributions from a weak $s$-process 
\citep{Timmes1995,Travaglio2004,venn2012apj}.
We note that our upper limit for Zr coincides with the lower error bound in HD 122563, while it should be produced in equal amounts to Y by the weak $r$-process. 
Ideally, an enhanced Sr abundance is expected from this process as well, but no Sr-feature fell within our spectral range. 

An alternative source of the excess in light $n$-capture elements is a 
weak $s$-process in fast rotating massive stars \citep{Frischknecht2016}. This scenario predicts an additional overproduction of O accompanying 
the enhancement in C, Sr,Y, and Zr.  Indeed, the [O/Fe] of 0.89 in ALW-8 is high, and also for HD 122563, a strongly elevated [O/Fe] ratio between 0.6 and 1.1 dex has
been reported \citep{Barbuy2003,Afsar2016}. Given the low number of elements that we measured in ALW-8 
beyond $Z\ga30$ , we cannot unambiguously 
conclude which of the processes ultimately led to the observed abundances and whether O was predominantly enhanced internally in the CNO processing
or already imprinted in the gas alongside with the primordial carbon from earlier generations of faint SNe and/or fast rotators. 
\section{Conclusions}
ALW-8 is the first CEMP-no star to be reported in the Carina dSph galaxy. 
The overall abundance pattern of the star suggests that the star was born from a medium that was enriched  by a mixture of 
faint SNe and/or fast rotating massive stars, and 
low-metallicity SNe II. The gas from which it formed had only undergone poor mixing. 
Moreover, the neutron-capture elements in  ALW-8 are  in excellent agreement with the only CEMP-no star in the Sculptor dSph and the weak-$r$ process star 
HD 122563. This suggests that, 
whichever process was responsible for the heavy element production must be an ubiquitous source to pollute the CEMP-no stars,  
acting independently of the environment, such as in the halo or in dSphs. 

While a wealth of CEMP-no stars is known in the 
MW halo, the absolute number of such stars identified in nearby dSph and ultra-faint dSph 
 satellites is low. 
Amongst these is the most iron-poor and most carbon-rich  star in the Segue~1 dSph  ([Fe/H] = $-$3.5, [C/Fe] = 2.3; \citealt{norris2010apj}).
While more iron-poor extragalactic stars exist in dSphs \citep{Frebel2010,Tafelmeyer2010AA}, none of these are carbon-rich.
On the other hand, the fraction of CEMP stars in the halo and in metal-poor dSph galaxies is known to significantly increase with decreasing metallicity, 
which has been interpreted as being due to the shift in the mean metallicity  from  lower   to higher values as the luminosity of the galaxy increases  \citep{salvadori2015mnras}.

Finally, the CEMP-no subclass is of particular interest as their fraction is largest in the outer Galactic halo, while 
CEMP-$s$ stars prevail in its inner, in-situ component \citep{Carollo2014}. This finding provides an immediate connection 
to the CEMP population in dSphs owing to the accretion origin of the outer halo \citep[e.g.,][]{Searle1978,Bullock2005,Carollo2007,Cooper2013}. As a consequence, 
the CEMP-no fraction in the Galactic satellites should be large 
and any additional candidate in this category is significant not only for nucleosynthetic considerations, but also 
for constraining the halo formation scenarios.
\begin{acknowledgements}
We thank M. Walker and T.T. Hansen for sharing their spectra of ALW-1 with us, and an anonymous referee for a helpful report.
A.S. and A.K. acknowledge the Deutsche Forschungsgemeinschaft for funding from  Emmy-Noether grant  Ko 4161/1. 
This work used data obtained from the INSPECT database, version 1.0 (www.inspect-stars.net).
 \end{acknowledgements}
\bibliographystyle{aa} 
\bibliography{reference_carina} 

\begin{thebibliography}{91}
\expandafter\ifx\csname natexlab\endcsname\relax\def\natexlab#1{#1}\fi

\bibitem[{{Abia} {et~al.}(2008){Abia}, {de Laverny}, \& {Wahlin}}]{abia2008aa}
{Abia}, C., {de Laverny}, P., \& {Wahlin}, R. 2008, \aap, 481, 161

\bibitem[{{Af{\c s}ar} {et~al.}(2016){Af{\c s}ar}, {Sneden}, {Frebel}, {Kim},
  {Mace}, {Kaplan}, {Lee}, {Oh}, {Sok Oh}, {Pak}, {Park}, {Pavel}, {Yuk}, \&
  {Jaffe}}]{Afsar2016}
{Af{\c s}ar}, M., {Sneden}, C., {Frebel}, A., {et~al.} 2016, \apj, 819, 103

\bibitem[{{Aoki} {et~al.}(2007){Aoki}, {Beers}, {Christlieb}, {Norris}, {Ryan},
  \& {Tsangarides}}]{aoki2007apj}
{Aoki}, W., {Beers}, T.~C., {Christlieb}, N., {et~al.} 2007, \apj, 655, 492

\bibitem[{{Arcones} \& {Montes}(2011)}]{Arcones2011}
{Arcones}, A. \& {Montes}, F. 2011, \apj, 731, 5

\bibitem[{{Asplund} {et~al.}(2009){Asplund}, {Grevesse}, {Sauval}, \&
  {Scott}}]{asplund2009araa}
{Asplund}, M., {Grevesse}, N., {Sauval}, A.~J., \& {Scott}, P. 2009, \araa, 47,
  481

\bibitem[{{Azzopardi} {et~al.}(1986){Azzopardi}, {Lequeux}, \&
  {Westerlund}}]{Azzopardi1986}
{Azzopardi}, M., {Lequeux}, J., \& {Westerlund}, B.~E. 1986, \aap, 161, 232

\bibitem[{{Barbuy} {et~al.}(2003){Barbuy}, {Mel{\'e}ndez}, {Spite}, {Spite},
  {Depagne}, {Hill}, {Cayrel}, {Bonifacio}, {Damineli}, \&
  {Torres}}]{Barbuy2003}
{Barbuy}, B., {Mel{\'e}ndez}, J., {Spite}, M., {et~al.} 2003, \apj, 588, 1072

\bibitem[{{Battaglia} {et~al.}(2008){Battaglia}, {Irwin}, {Tolstoy}, {Hill},
  {Helmi}, {Letarte}, \& {Jablonka}}]{battaglia2008mnras}
{Battaglia}, G., {Irwin}, M., {Tolstoy}, E., {et~al.} 2008, \mnras, 383, 183

\bibitem[{{Battistini} \& {Bensby}(2016)}]{Battistini2016}
{Battistini}, C. \& {Bensby}, T. 2016, \aap, 586, A49

\bibitem[{{Beers} \& {Christlieb}(2005)}]{beers-christlieb2005ARA&A}
{Beers}, T.~C. \& {Christlieb}, N. 2005, \araa, 43, 531

\bibitem[{{Bensby} {et~al.}(2014){Bensby}, {Feltzing}, \& {Oey}}]{Bensby2014}
{Bensby}, T., {Feltzing}, S., \& {Oey}, M.~S. 2014, \aap, 562, A71

\bibitem[{{Bonifacio} {et~al.}(2015){Bonifacio}, {Caffau}, {Spite}, {Limongi},
  {Chieffi}, {Klessen}, {Fran{\c c}ois}, {Molaro}, {Ludwig}, {Zaggia}, {Spite},
  {Plez}, {Cayrel}, {Christlieb}, {Clark}, {Glover}, {Hammer}, {Koch},
  {Monaco}, {Sbordone}, \& {Steffen}}]{bonifacio2015aa}
{Bonifacio}, P., {Caffau}, E., {Spite}, M., {et~al.} 2015, \aap, 579, A28

\bibitem[{{Bullock} \& {Johnston}(2005)}]{Bullock2005}
{Bullock}, J.~S. \& {Johnston}, K.~V. 2005, \apj, 635, 931

\bibitem[{{Carollo} {et~al.}(2007){Carollo}, {Beers}, {Lee}, {Chiba}, {Norris},
  {Wilhelm}, {Sivarani}, {Marsteller}, {Munn}, {Bailer-Jones}, {Fiorentin}, \&
  {York}}]{Carollo2007}
{Carollo}, D., {Beers}, T.~C., {Lee}, Y.~S., {et~al.} 2007, \nat, 450, 1020

\bibitem[{{Carollo} {et~al.}(2014){Carollo}, {Freeman}, {Beers}, {Placco},
  {Tumlinson}, \& {Martell}}]{Carollo2014}
{Carollo}, D., {Freeman}, K., {Beers}, T.~C., {et~al.} 2014, \apj, 788, 180

\bibitem[{{Cayrel de Strobel} \& {Spite}(1988)}]{cayrel1988iaus}
{Cayrel de Strobel}, G. \& {Spite}, M., eds. 1988, IAU Symposium, Vol. 132,
  {The impact of very high S/N spectroscopy on stellar physics: proceedings of
  the 132nd Symposium of the International Astronomical Union held in Paris,
  France, June 29-July 3, 1987.}

\bibitem[{{Charbonnel} {et~al.}(1998){Charbonnel}, {Brown}, \&
  {Wallerstein}}]{charbonnel1998aa}
{Charbonnel}, C., {Brown}, J.~A., \& {Wallerstein}, G. 1998, \aap, 332, 204

\bibitem[{{Cooper} {et~al.}(2013){Cooper}, {D'Souza}, {Kauffmann}, {Wang},
  {Boylan-Kolchin}, {Guo}, {Frenk}, \& {White}}]{Cooper2013}
{Cooper}, A.~P., {D'Souza}, R., {Kauffmann}, G., {et~al.} 2013, \mnras, 434,
  3348

\bibitem[{{Cutri} {et~al.}(2003){Cutri}, {Skrutskie}, {van Dyk}, {Beichman},
  {Carpenter}, {Chester}, {Cambresy}, {Evans}, {Fowler}, {Gizis}, {Howard},
  {Huchra}, {Jarrett}, {Kopan}, {Kirkpatrick}, {Light}, {Marsh}, {McCallon},
  {Schneider}, {Stiening}, {Sykes}, {Weinberg}, {Wheaton}, {Wheelock}, \&
  {Zacarias}}]{Cutri2003}
{Cutri}, R.~M., {Skrutskie}, M.~F., {van Dyk}, S., {et~al.} 2003, VizieR Online
  Data Catalog, 2246, 0

\bibitem[{{Fabrizio} {et~al.}(2015){Fabrizio}, {Nonino}, {Bono}, {Primas},
  {Th{\'e}venin}, {Stetson}, {Cassisi}, {Buonanno}, {Coppola}, {da Silva},
  {Dall'Ora}, {Ferraro}, {Genovali}, {Gilmozzi}, {Iannicola}, {Marconi},
  {Monelli}, {Romaniello}, \& {Walker}}]{fabrizio2015aa}
{Fabrizio}, M., {Nonino}, M., {Bono}, G., {et~al.} 2015, \aap, 580, A18

\bibitem[{{Frebel} {et~al.}(2010{\natexlab{a}}){Frebel}, {Kirby}, \&
  {Simon}}]{Frebel2010}
{Frebel}, A., {Kirby}, E.~N., \& {Simon}, J.~D. 2010{\natexlab{a}}, \nat, 464,
  72

\bibitem[{{Frebel} {et~al.}(2010{\natexlab{b}}){Frebel}, {Simon}, {Geha}, \&
  {Willman}}]{frebel2010apj}
{Frebel}, A., {Simon}, J.~D., {Geha}, M., \& {Willman}, B. 2010{\natexlab{b}},
  \apj, 708, 560

\bibitem[{{Frischknecht} {et~al.}(2016){Frischknecht}, {Hirschi}, {Pignatari},
  {Maeder}, {Meynet}, {Chiappini}, {Thielemann}, {Rauscher}, {Georgy}, \&
  {Ekstr{\"o}m}}]{Frischknecht2016}
{Frischknecht}, U., {Hirschi}, R., {Pignatari}, M., {et~al.} 2016, \mnras, 456,
  1803

\bibitem[{{Geisler} {et~al.}(2005){Geisler}, {Smith}, {Wallerstein},
  {Gonzalez}, \& {Charbonnel}}]{Geisler2005}
{Geisler}, D., {Smith}, V.~V., {Wallerstein}, G., {Gonzalez}, G., \&
  {Charbonnel}, C. 2005, \aj, 129, 1428

\bibitem[{{Geisler} {et~al.}(2007){Geisler}, {Wallerstein}, {Smith}, \&
  {Casetti-Dinescu}}]{Geisler2007}
{Geisler}, D., {Wallerstein}, G., {Smith}, V.~V., \& {Casetti-Dinescu}, D.~I.
  2007, \pasp, 119, 939

\bibitem[{{Gratton} {et~al.}(2000){Gratton}, {Sneden}, {Carretta}, \&
  {Bragaglia}}]{gratton2000aa}
{Gratton}, R.~G., {Sneden}, C., {Carretta}, E., \& {Bragaglia}, A. 2000, \aap,
  354, 169

\bibitem[{{Grebel}(1997)}]{Grebel1997}
{Grebel}, E.~K. 1997, in Reviews in Modern Astronomy, Vol.~10, Reviews in
  Modern Astronomy, ed. R.~E. {Schielicke}, 29--60

\bibitem[{{Hanke} {et~al.}(2017){Hanke}, {Koch}, {Hansen}, \&
  {McWilliam}}]{Hanke2017}
{Hanke}, M., {Koch}, A., {Hansen}, C.~J., \& {McWilliam}, A. 2017, \aap, 599,
  97

\bibitem[{{Hansen} {et~al.}(2011){Hansen}, {Nordstr{\"o}m}, {Bonifacio},
  {Spite}, {Andersen}, {Beers}, {Cayrel}, {Spite}, {Molaro}, {Barbuy},
  {Depagne}, {Fran{\c c}ois}, {Hill}, {Plez}, \& {Sivarani}}]{CJHansen2011}
{Hansen}, C.~J., {Nordstr{\"o}m}, B., {Bonifacio}, P., {et~al.} 2011, \aap,
  527, A65

\bibitem[{{Hansen} {et~al.}(2016{\natexlab{a}}){Hansen}, {Nordstr{\"o}m},
  {Hansen}, {Kennedy}, {Placco}, {Beers}, {Andersen}, {Cescutti}, \&
  {Chiappini}}]{cjhansen2016}
{Hansen}, C.~J., {Nordstr{\"o}m}, B., {Hansen}, T.~T., {et~al.}
  2016{\natexlab{a}}, \aap, 588, A37

\bibitem[{{Hansen} {et~al.}(2016{\natexlab{b}}){Hansen}, {Andersen},
  {Nordstr{\"o}m}, {Beers}, {Placco}, {Yoon}, \& {Buchhave}}]{hansentt2016aa}
{Hansen}, T.~T., {Andersen}, J., {Nordstr{\"o}m}, B., {et~al.}
  2016{\natexlab{b}}, \aap, 586, A160

\bibitem[{{Hansen} {et~al.}(2016{\natexlab{c}}){Hansen}, {Andersen},
  {Nordstr{\"o}m}, {Beers}, {Placco}, {Yoon}, \& {Buchhave}}]{hansentt2016aa2}
{Hansen}, T.~T., {Andersen}, J., {Nordstr{\"o}m}, B., {et~al.}
  2016{\natexlab{c}}, \aap, 588, A3

\bibitem[{{Honda} {et~al.}(2011){Honda}, {Aoki}, {Arimoto}, \&
  {Sadakane}}]{honda2011pasj}
{Honda}, S., {Aoki}, W., {Arimoto}, N., \& {Sadakane}, K. 2011, \pasj, 63, 523

\bibitem[{{Honda} {et~al.}(2006){Honda}, {Aoki}, {Ishimaru}, {Wanajo}, \&
  {Ryan}}]{Honda2006}
{Honda}, S., {Aoki}, W., {Ishimaru}, Y., {Wanajo}, S., \& {Ryan}, S.~G. 2006,
  \apj, 643, 1180

\bibitem[{{Idiart} \& {Th{\'e}venin}(2000)}]{idiart2000apj}
{Idiart}, T. \& {Th{\'e}venin}, F. 2000, \apj, 541, 207

\bibitem[{{Ishigaki} {et~al.}(2014){Ishigaki}, {Tominaga}, {Kobayashi}, \&
  {Nomoto}}]{Ishigaki2014}
{Ishigaki}, M.~N., {Tominaga}, N., {Kobayashi}, C., \& {Nomoto}, K. 2014,
  \apjl, 792, L32

\bibitem[{{Ji} {et~al.}(2016{\natexlab{a}}){Ji}, {Frebel}, {Ezzeddine}, \&
  {Casey}}]{tuc-ji2016apj}
{Ji}, A.~P., {Frebel}, A., {Ezzeddine}, R., \& {Casey}, A.~R.
  2016{\natexlab{a}}, \apjl, 832, L3

\bibitem[{{Ji} {et~al.}(2016{\natexlab{b}}){Ji}, {Frebel}, {Simon}, \&
  {Chiti}}]{ret-ji2016apj}
{Ji}, A.~P., {Frebel}, A., {Simon}, J.~D., \& {Chiti}, A. 2016{\natexlab{b}},
  \apj, 830, 93

\bibitem[{{Kirby} {et~al.}(2011){Kirby}, {Lanfranchi}, {Simon}, {Cohen}, \&
  {Guhathakurta}}]{Kirby2011}
{Kirby}, E.~N., {Lanfranchi}, G.~A., {Simon}, J.~D., {Cohen}, J.~G., \&
  {Guhathakurta}, P. 2011, \apj, 727, 78

\bibitem[{{Kobayashi} {et~al.}(2011){Kobayashi}, {Tominaga}, \&
  {Nomoto}}]{Kobayashi2011}
{Kobayashi}, C., {Tominaga}, N., \& {Nomoto}, K. 2011, \apjl, 730, L14

\bibitem[{{Kobayashi} {et~al.}(2006){Kobayashi}, {Umeda}, {Nomoto}, {Tominaga},
  \& {Ohkubo}}]{Kobayashi2006ApJ}
{Kobayashi}, C., {Umeda}, H., {Nomoto}, K., {Tominaga}, N., \& {Ohkubo}, T.
  2006, \apj, 653, 1145

\bibitem[{{Koch} \& {Edvardsson}(2002)}]{Koch2002}
{Koch}, A. \& {Edvardsson}, B. 2002, \aap, 381, 500

\bibitem[{{Koch} {et~al.}(2008){Koch}, {Grebel}, {Gilmore}, {Wyse}, {Kleyna},
  {Harbeck}, {Wilkinson}, \& {Wyn Evans}}]{koch2008aj}
{Koch}, A., {Grebel}, E.~K., {Gilmore}, G.~F., {et~al.} 2008, \aj, 135, 1580

\bibitem[{{Koch} {et~al.}(2006){Koch}, {Grebel}, {Wyse}, {Kleyna}, {Wilkinson},
  {Harbeck}, {Gilmore}, \& {Evans}}]{koch2006aj}
{Koch}, A., {Grebel}, E.~K., {Wyse}, R.~F.~G., {et~al.} 2006, \aj, 131, 895

\bibitem[{{Koch} {et~al.}(2016){Koch}, {McWilliam}, {Preston}, \&
  {Thompson}}]{Koch2016}
{Koch}, A., {McWilliam}, A., {Preston}, G.~W., \& {Thompson}, I.~B. 2016, \aap,
  587, A124

\bibitem[{{Kupka} {et~al.}(1999){Kupka}, {Piskunov}, {Ryabchikova}, {Stempels},
  \& {Weiss}}]{kupka1999aas}
{Kupka}, F., {Piskunov}, N., {Ryabchikova}, T.~A., {Stempels}, H.~C., \&
  {Weiss}, W.~W. 1999, \aaps, 138, 119

\bibitem[{{Lai} {et~al.}(2011){Lai}, {Lee}, {Bolte}, {Lucatello}, {Beers},
  {Johnson}, {Sivarani}, \& {Rockosi}}]{lai2011apj}
{Lai}, D.~K., {Lee}, Y.~S., {Bolte}, M., {et~al.} 2011, \apj, 738, 51

\bibitem[{{Lee} {et~al.}(2017){Lee}, {Beers}, {Kim}, {Placco}, {Yoon},
  {Carollo}, {Masseron}, \& {Jung}}]{Lee2017}
{Lee}, Y.~S., {Beers}, T.~C., {Kim}, Y.~K., {et~al.} 2017, \apj, 836, 91

\bibitem[{{Lemasle} {et~al.}(2012){Lemasle}, {Hill}, {Tolstoy}, {Venn},
  {Shetrone}, {Irwin}, {de Boer}, {Starkenburg}, \&
  {Salvadori}}]{lemasle2012aa}
{Lemasle}, B., {Hill}, V., {Tolstoy}, E., {et~al.} 2012, \aap, 538, A100

\bibitem[{{Lind} {et~al.}(2011){Lind}, {Asplund}, {Barklem}, \&
  {Belyaev}}]{Lind2011}
{Lind}, K., {Asplund}, M., {Barklem}, P.~S., \& {Belyaev}, A.~K. 2011, \aap,
  528, A103

\bibitem[{{Lind} {et~al.}(2009){Lind}, {Primas}, {Charbonnel}, {Grundahl}, \&
  {Asplund}}]{lind2009}
{Lind}, K., {Primas}, F., {Charbonnel}, C., {Grundahl}, F., \& {Asplund}, M.
  2009, \aap, 503, 545

\bibitem[{{Masseron} {et~al.}(2012){Masseron}, {Johnson}, {Lucatello},
  {Karakas}, {Plez}, {Beers}, \& {Christlieb}}]{Masseron2012}
{Masseron}, T., {Johnson}, J.~A., {Lucatello}, S., {et~al.} 2012, \apj, 751, 14

\bibitem[{{Masseron} {et~al.}(2010){Masseron}, {Johnson}, {Plez}, {van Eck},
  {Primas}, {Goriely}, \& {Jorissen}}]{masseron2010aa}
{Masseron}, T., {Johnson}, J.~A., {Plez}, B., {et~al.} 2010, \aap, 509, A93

\bibitem[{{M{\'e}sz{\'a}ros} {et~al.}(2012){M{\'e}sz{\'a}ros}, {Allende
  Prieto}, {Edvardsson}, {Castelli}, {Garc{\'{\i}}a P{\'e}rez}, {Gustafsson},
  {Majewski}, {Plez}, {Schiavon}, {Shetrone}, \& {de Vicente}}]{meszaros2012aj}
{M{\'e}sz{\'a}ros}, S., {Allende Prieto}, C., {Edvardsson}, B., {et~al.} 2012,
  \aj, 144, 120

\bibitem[{{Meynet} {et~al.}(2006){Meynet}, {Ekstr{\"o}m}, \&
  {Maeder}}]{Meynet2006}
{Meynet}, G., {Ekstr{\"o}m}, S., \& {Maeder}, A. 2006, \aap, 447, 623

\bibitem[{{Monelli} {et~al.}(2003){Monelli}, {Pulone}, {Corsi}, {Castellani},
  {Bono}, {Walker}, {Brocato}, {Buonanno}, {Caputo}, {Castellani}, {Dall'Ora},
  {Marconi}, {Nonino}, {Ripepi}, \& {Smith}}]{Monelli2003}
{Monelli}, M., {Pulone}, L., {Corsi}, C.~E., {et~al.} 2003, \aj, 126, 218

\bibitem[{{Mould} \& {Aaronson}(1983)}]{Mould1983}
{Mould}, J. \& {Aaronson}, M. 1983, \apj, 273, 530

\bibitem[{{Nonino} {et~al.}(1999){Nonino}, {Bertin}, {da Costa}, {Deul},
  {Erben}, {Olsen}, {Prandoni}, {Scodeggio}, {Wicenec}, {Wichmann}, {Benoist},
  {Freudling}, {Guarnieri}, {Hook}, {Hook}, {Mendez}, {Savaglio}, {Silva}, \&
  {Slijkhuis}}]{nonino1999aas}
{Nonino}, M., {Bertin}, E., {da Costa}, L., {et~al.} 1999, \aaps, 137, 51

\bibitem[{{Norris} {et~al.}(2010{\natexlab{a}}){Norris}, {Gilmore}, {Wyse},
  {Yong}, \& {Frebel}}]{norris2010apj2}
{Norris}, J.~E., {Gilmore}, G., {Wyse}, R.~F.~G., {Yong}, D., \& {Frebel}, A.
  2010{\natexlab{a}}, \apjl, 722, L104

\bibitem[{{Norris} {et~al.}(2010{\natexlab{b}}){Norris}, {Wyse}, {Gilmore},
  {Yong}, {Frebel}, {Wilkinson}, {Belokurov}, \& {Zucker}}]{norris2010apj}
{Norris}, J.~E., {Wyse}, R.~F.~G., {Gilmore}, G., {et~al.} 2010{\natexlab{b}},
  \apj, 723, 1632

\bibitem[{{Placco} {et~al.}(2014){Placco}, {Frebel}, {Beers}, \&
  {Stancliffe}}]{placco2014apj}
{Placco}, V.~M., {Frebel}, A., {Beers}, T.~C., \& {Stancliffe}, R.~J. 2014,
  \apj, 797, 21

\bibitem[{{Plez}(2012)}]{Plez2012ascl}
{Plez}, B. 2012, {Turbospectrum: Code for spectral synthesis}, Astrophysics
  Source Code Library

\bibitem[{{Plez} \& {Cohen}(2005)}]{plez-cohen2005aa}
{Plez}, B. \& {Cohen}, J.~G. 2005, \aap, 434, 1117

\bibitem[{{Roederer} {et~al.}(2014){Roederer}, {Preston}, {Thompson},
  {Shectman}, {Sneden}, {Burley}, \& {Kelson}}]{Roederer2014}
{Roederer}, I.~U., {Preston}, G.~W., {Thompson}, I.~B., {et~al.} 2014, \aj,
  147, 136

\bibitem[{{Salgado} {et~al.}(2016){Salgado}, {Da Costa}, {Yong}, \&
  {Norris}}]{salgado2016mnras}
{Salgado}, C., {Da Costa}, G.~S., {Yong}, D., \& {Norris}, J.~E. 2016, \mnras,
  463, 598

\bibitem[{{Salvadori} {et~al.}(2015){Salvadori}, {Sk{\'u}lad{\'o}ttir}, \&
  {Tolstoy}}]{salvadori2015mnras}
{Salvadori}, S., {Sk{\'u}lad{\'o}ttir}, {\'A}., \& {Tolstoy}, E. 2015, \mnras,
  454, 1320

\bibitem[{{Sbordone} {et~al.}(2010){Sbordone}, {Bonifacio}, {Caffau}, {Ludwig},
  {Behara}, {Gonz{\'a}lez Hern{\'a}ndez}, {Steffen}, {Cayrel}, {Freytag},
  {van't Veer}, {Molaro}, {Plez}, {Sivarani}, {Spite}, {Spite}, {Beers},
  {Christlieb}, {Fran{\c c}ois}, \& {Hill}}]{Sbordone2010}
{Sbordone}, L., {Bonifacio}, P., {Caffau}, E., {et~al.} 2010, \aap, 522, A26

\bibitem[{{Schlegel} {et~al.}(1998){Schlegel}, {Finkbeiner}, \&
  {Davis}}]{Schlegel1998}
{Schlegel}, D.~J., {Finkbeiner}, D.~P., \& {Davis}, M. 1998, \apj, 500, 525

\bibitem[{{Searle} \& {Zinn}(1978)}]{Searle1978}
{Searle}, L. \& {Zinn}, R. 1978, \apj, 225, 357

\bibitem[{{Shetrone} {et~al.}(2003){Shetrone}, {Venn}, {Tolstoy}, {Primas},
  {Hill}, \& {Kaufer}}]{shetrone2003aj}
{Shetrone}, M., {Venn}, K.~A., {Tolstoy}, E., {et~al.} 2003, \aj, 125, 684

\bibitem[{{Simmerer} {et~al.}(2004){Simmerer}, {Sneden}, {Cowan}, {Collier},
  {Woolf}, \& {Lawler}}]{Simmerer2004}
{Simmerer}, J., {Sneden}, C., {Cowan}, J.~J., {et~al.} 2004, \apj, 617, 1091

\bibitem[{{Sivarani} {et~al.}(2006){Sivarani}, {Beers}, {Bonifacio}, {Molaro},
  {Cayrel}, {Herwig}, {Spite}, {Spite}, {Plez}, {Andersen}, {Barbuy},
  {Depagne}, {Hill}, {Fran{\c c}ois}, {Nordstr{\"o}m}, \&
  {Primas}}]{sivarani2006aa}
{Sivarani}, T., {Beers}, T.~C., {Bonifacio}, P., {et~al.} 2006, \aap, 459, 125

\bibitem[{{Sk{\'u}lad{\'o}ttir} {et~al.}(2015){Sk{\'u}lad{\'o}ttir}, {Tolstoy},
  {Salvadori}, {Hill}, {Pettini}, {Shetrone}, \&
  {Starkenburg}}]{skuladottir2015aa}
{Sk{\'u}lad{\'o}ttir}, {\'A}., {Tolstoy}, E., {Salvadori}, S., {et~al.} 2015,
  \aap, 574, A129

\bibitem[{{Smecker-Hane} {et~al.}(1994){Smecker-Hane}, {Stetson}, {Hesser}, \&
  {Lehnert}}]{Smecker-Hane1994}
{Smecker-Hane}, T.~A., {Stetson}, P.~B., {Hesser}, J.~E., \& {Lehnert}, M.~D.
  1994, \aj, 108, 507

\bibitem[{{Sneden} {et~al.}(2008){Sneden}, {Cowan}, \&
  {Gallino}}]{sneden2008ARAA}
{Sneden}, C., {Cowan}, J.~J., \& {Gallino}, R. 2008, \araa, 46, 241

\bibitem[{{Sneden} {et~al.}(2003){Sneden}, {Cowan}, {Lawler}, {Ivans},
  {Burles}, {Beers}, {Primas}, {Hill}, {Truran}, {Fuller}, {Pfeiffer}, \&
  {Kratz}}]{Sneden2003}
{Sneden}, C., {Cowan}, J.~J., {Lawler}, J.~E., {et~al.} 2003, \apj, 591, 936

\bibitem[{{Spite} {et~al.}(2006){Spite}, {Cayrel}, {Hill}, {Spite}, {Fran{\c
  c}ois}, {Plez}, {Bonifacio}, {Molaro}, {Depagne}, {Andersen}, {Barbuy},
  {Beers}, {Nordstr{\"o}m}, \& {Primas}}]{Spite2006A&A}
{Spite}, M., {Cayrel}, R., {Hill}, V., {et~al.} 2006, \aap, 455, 291

\bibitem[{{Spite} {et~al.}(2005){Spite}, {Cayrel}, {Plez}, {Hill}, {Spite},
  {Depagne}, {Fran{\c c}ois}, {Bonifacio}, {Barbuy}, {Beers}, {Andersen},
  {Molaro}, {Nordstr{\"o}m}, \& {Primas}}]{Spite2005}
{Spite}, M., {Cayrel}, R., {Plez}, B., {et~al.} 2005, \aap, 430, 655

\bibitem[{{Spite} \& {Spite}(1982)}]{Spite1982}
{Spite}, M. \& {Spite}, F. 1982, \nat, 297, 483

\bibitem[{{Starkenburg} {et~al.}(2014){Starkenburg}, {Shetrone}, {McConnachie},
  \& {Venn}}]{Starkenburg2014}
{Starkenburg}, E., {Shetrone}, M.~D., {McConnachie}, A.~W., \& {Venn}, K.~A.
  2014, \mnras, 441, 1217

\bibitem[{{Tafelmeyer} {et~al.}(2010){Tafelmeyer}, {Jablonka}, {Hill},
  {Shetrone}, {Tolstoy}, {Irwin}, {Battaglia}, {Helmi}, {Starkenburg}, {Venn},
  {Abel}, {Francois}, {Kaufer}, {North}, {Primas}, \&
  {Szeifert}}]{Tafelmeyer2010AA}
{Tafelmeyer}, M., {Jablonka}, P., {Hill}, V., {et~al.} 2010, \aap, 524, A58

\bibitem[{{Thompson} {et~al.}(2008){Thompson}, {Ivans}, {Bisterzo}, {Sneden},
  {Gallino}, {Vauclair}, {Burley}, {Shectman}, \& {Preston}}]{thompson2008apj}
{Thompson}, I.~B., {Ivans}, I.~I., {Bisterzo}, S., {et~al.} 2008, \apj, 677,
  556

\bibitem[{{Timmes} {et~al.}(1995){Timmes}, {Woosley}, \& {Weaver}}]{Timmes1995}
{Timmes}, F.~X., {Woosley}, S.~E., \& {Weaver}, T.~A. 1995, \apjs, 98, 617

\bibitem[{{Tolstoy} {et~al.}(2009){Tolstoy}, {Hill}, \& {Tosi}}]{Tolstoy2009}
{Tolstoy}, E., {Hill}, V., \& {Tosi}, M. 2009, \araa, 47, 371

\bibitem[{{Tolstoy} {et~al.}(2003){Tolstoy}, {Venn}, {Shetrone}, {Primas},
  {Hill}, {Kaufer}, \& {Szeifert}}]{Tolstoy2003}
{Tolstoy}, E., {Venn}, K.~A., {Shetrone}, M., {et~al.} 2003, \aj, 125, 707

\bibitem[{{Tominaga} {et~al.}(2007){Tominaga}, {Umeda}, \&
  {Nomoto}}]{Tominaga2007}
{Tominaga}, N., {Umeda}, H., \& {Nomoto}, K. 2007, \apj, 660, 516

\bibitem[{{Travaglio} {et~al.}(2004){Travaglio}, {Gallino}, {Arnone}, {Cowan},
  {Jordan}, \& {Sneden}}]{Travaglio2004}
{Travaglio}, C., {Gallino}, R., {Arnone}, E., {et~al.} 2004, \apj, 601, 864

\bibitem[{{Umeda} \& {Nomoto}(2003)}]{Umeda2003}
{Umeda}, H. \& {Nomoto}, K. 2003, \nat, 422, 871

\bibitem[{{Venn} {et~al.}(2012){Venn}, {Shetrone}, {Irwin}, {Hill}, {Jablonka},
  {Tolstoy}, {Lemasle}, {Divell}, {Starkenburg}, {Letarte}, {Baldner},
  {Battaglia}, {Helmi}, {Kaufer}, \& {Primas}}]{venn2012apj}
{Venn}, K.~A., {Shetrone}, M.~D., {Irwin}, M.~J., {et~al.} 2012, \apj, 751, 102

\bibitem[{{Walker} {et~al.}(2009){Walker}, {Mateo}, \&
  {Olszewski}}]{Walker2009}
{Walker}, M.~G., {Mateo}, M., \& {Olszewski}, E.~W. 2009, \aj, 137, 3100

\bibitem[{{Yoon} {et~al.}(2016){Yoon}, {Beers}, {Placco}, {Rasmussen},
  {Carollo}, {He}, {Hansen}, {Roederer}, \& {Zeanah}}]{Yoon2016}
{Yoon}, J., {Beers}, T.~C., {Placco}, V.~M., {et~al.} 2016, \apj, 833, 20

\end{thebibliography}

\onecolumn
\begin{appendix}
\begin{table}
\caption{Line list for the EW analysis, abundance and abundance uncertainties to 1 $\sigma$ scatter in the 
stellar parameters for ALW-8.}
\centering
\begin{tabular}{cccrrrccc}
\hline\hline
Element&$\lambda$& $\chi$  & log(gf) &   EW     & A(x)   &$\delta$ T    & $\delta$ log $\textit{g}$  & $ \delta \xi $\\
       & \AA\    &  (eV)   &        &    m\AA\   &  &$\pm$100 K &  $\pm$0.25 dex    & $\pm$0.2 km\,s$^{-1}$ \\

\hline
Li I   & 6707.70  & 0.00 &      0.17  &  syn   &     0.00   &   $\pm$0.30    &  $\mp$0.10    &  $\pm$0.01 \\
O I    & 6300.31  & 0.00 &   $-$9.82  &  syn   & 6.93   &   $\mp$0.10   &  $\mp$0.03   &  $\pm$0.01 \\
O I    & 6363.78  & 0.02 &  $-$10.30  &  syn   & 7.23   &      $\mp$0.10   &  $\mp$0.03  &  $\pm$0.01 \\
Na  I  & 5682.63  & 2.09 &   $-$0.70  &   58   &  4.84  &   $\mp$0.03   &  $\pm$0.07   &  $\mp$0.02 \\
Na  I  & 5688.21  & 2.09 &   $-$0.45  &   82   &  4.91  &   $\mp$0.02   &  $\pm$0.09   &  $\mp$0.03 \\
Mg  I  & 5528.40  & 4.34 &   $-$0.62  &  120   &  5.57  &   $\mp$0.00   &  $\pm$0.11   &  $\mp$0.06 \\
Ca  I  & 5261.70  & 2.52 &   $-$0.59  &   67   &  4.51  &   $\mp$0.04   &  $\pm$0.08   &  $\mp$0.04 \\
Ca  I  & 5588.75  & 2.53 &      0.31  &  100   &  4.07  &   $\mp$0.02   &  $\pm$0.10   &  $\mp$0.07 \\
Ca  I  & 5594.46  & 2.52 &      0.05  &  102   &  4.36  &   $\mp$0.02   &  $\pm$0.10   &  $\mp$0.08 \\
Ca  I  & 6102.72  & 1.88 &   $-$0.86  &   85   &  4.05  &   $\mp$0.06   &  $\pm$0.09   &  $\mp$0.04 \\
Ca  I  & 6122.22  & 1.89 &   $-$0.38  &  167   &  4.66  &   $\mp$0.05   &  $\pm$0.15   &  $\mp$0.09 \\
Sc II  & 5526.79  & 1.77 &      0.02  &  100   &  1.27  &   $\pm$0.00   &  $\mp$0.13   &  $\mp$0.08 \\
Sc II  & 5657.90  & 1.51 &   $-$0.60  &   70   &  1.08  &   $\mp$0.02   &  $\mp$0.16   &  $\mp$0.04 \\
Sc II  & 5684.20  & 1.51 &   $-$1.07  &   55   &  1.15  &   $\mp$0.03   &  $\mp$0.16   &  $\mp$0.02 \\
Ti  I  & 4981.73  & 0.85 &      0.50  &  180   &  3.26  &   $\mp$0.11   &  $\pm$0.14   &  $\mp$0.15 \\
Ti  I  & 4999.50  & 0.83 &      0.25  &   95   &  2.14  &   $\mp$0.11   &  $\pm$0.10   &  $\mp$0.07 \\
Ti  I  & 5020.03  & 0.84 &   $-$0.41  &   84   &  2.66  &   $\mp$0.12   &  $\pm$0.10   &  $\mp$0.05 \\
Ti  I  & 5024.84  & 0.82 &   $-$0.60  &   47   &  2.30  &   $\mp$0.14   &  $\pm$0.08   &  $\mp$0.02 \\
Ti  I  & 5036.46  & 1.44 &      0.13  &   60   &  2.63  &   $\mp$0.11   &  $\pm$0.09   &  $\mp$0.03 \\
Ti  I  & 5173.74  & 0.00 &   $-$1.11  &  135   &  2.83  &   $\mp$0.15   &  $\pm$0.08   &  $\mp$0.12 \\
Ti  I  & 5210.38  & 0.05 &   $-$0.88  &  120   &  2.44  &   $\mp$0.15   &  $\pm$0.09   &  $\mp$0.09 \\
Ti II  & 4865.61  & 1.12 &   $-$2.67  &   51   &  2.78  &   $\mp$0.04   &  $\mp$0.15   &  $\mp$0.02 \\
Ti II  & 5336.77  & 1.58 &   $-$1.63  &   97   &  3.01  &   $\mp$0.00   &  $\mp$0.14   &  $\mp$0.08 \\
Ti II  & 6491.56  & 2.05 &   $-$1.79  &   33   &  2.74  &   $\mp$0.02   &  $\mp$0.17   &  $\mp$0.01 \\
Cr  I  & 5247.57  & 0.96 &   $-$1.64  &   66   &  2.99  &   $\mp$0.12   &  $\pm$0.09   &  $\mp$0.03 \\
Cr  I  & 5296.69  & 0.98 &   $-$1.40  &   95   &  3.18  &   $\mp$0.11   &  $\pm$0.10   &  $\mp$0.07 \\
Cr  I  & 5345.80  & 1.00 &   $-$0.98  &  139   &  3.44  &   $\mp$0.10   &  $\pm$0.11   &  $\mp$0.12 \\
Cr  I  & 5348.31  & 1.00 &   $-$1.29  &   64   &  2.66  &   $\mp$0.12   &  $\pm$0.09   &  $\mp$0.03 \\
Cr  I  & 5409.77  & 1.03 &   $-$0.72  &  120   &  2.91  &   $\mp$0.10   &  $\pm$0.10   &  $\mp$0.10 \\
Mn  I  & 5394.68  & 0.00 &   $-$3.50  &   70   &  2.61  &   $\mp$0.18   &  $\pm$0.06   &  $\mp$0.03 \\
Mn  I  & 5432.55  & 0.00 &   $-$3.79  &   54   &  2.70  &   $\mp$0.18   &  $\pm$0.06   &  $\mp$0.02 \\
Fe  I  & 4891.49  & 2.85 &   $-$0.11  &  160   &  4.81  &   $\mp$0.04   &  $\pm$0.14   &  $\mp$0.11 \\
Fe  I  & 4966.09  & 3.32 &   $-$0.87  &   94   &  5.22  &   $\mp$0.04   &  $\pm$0.07   &  $\mp$0.08 \\
Fe  I  & 5001.86  & 3.88 &      0.01  &   88   &  5.01  &   $\mp$0.02   &  $\pm$0.06   &  $\mp$0.07 \\
Fe  I  & 5194.94  & 1.55 &   $-$2.08  &  156   &  4.90  &   $\mp$0.09   &  $\pm$0.06   &  $\mp$0.16 \\
Fe  I  & 5195.47  & 4.21 &   $-$0.08  &   57   &  5.05  &   $\mp$0.04   &  $\pm$0.04   &  $\mp$0.03 \\
Fe  I  & 5215.18  & 3.27 &   $-$0.87  &  100   &  5.17  &   $\mp$0.04   &  $\pm$0.07   &  $\mp$0.08 \\
Fe  I  & 5225.53  & 0.11 &   $-$4.78  &  128   &  4.97  &   $\mp$0.15   &  $\pm$0.01   &  $\mp$0.12 \\
Fe  I  & 5242.49  & 3.63 &   $-$0.96  &   50   &  5.04  &   $\mp$0.06   &  $\pm$0.03   &  $\mp$0.02 \\
Fe  I  & 5307.36  & 1.61 &   $-$2.98  &   93   &  4.84  &   $\mp$0.11   &  $\pm$0.04   &  $\mp$0.07 \\
Fe  I  & 5586.76  & 3.36 &   $-$0.12  &  133   &  4.97  &   $\mp$0.03   &  $\pm$0.09   &  $\mp$0.10 \\
Fe  I  & 6546.24  & 2.76 &   $-$1.53  &  103   &  4.97  &   $\mp$0.08   &  $\pm$0.02   &  $\mp$0.07 \\
Fe II  & 5197.58  & 3.23 &   $-$2.09  &   55   &  5.03  &   $\pm$0.05   &  $\mp$0.18   &  $\mp$0.03 \\
Fe II  & 5234.63  & 3.22 &   $-$2.23  &   45   &  4.96  &   $\pm$0.04   &  $\mp$0.18   &  $\mp$0.02 \\
Ni  I  & 4980.17  & 3.60 &      0.00  &   95   &  4.62  &   $\mp$0.01   &  $\pm$0.04   &  $\mp$0.08 \\
Ni  I  & 5424.64  & 1.95 &   $-$2.77  &   85   &  4.84  &   $\mp$0.08   &  $\pm$0.00   &  $\mp$0.06 \\
Ni  I  & 6108.11  & 1.67 &   $-$2.45  &   52   &  3.59  &   $\mp$0.11   &  $\mp$0.01   &  $\mp$0.02 \\
Ni  I  & 6314.65  & 1.92 &   $-$1.77  &   62   &  3.39  &   $\mp$0.10   &  $\mp$0.01   &  $\mp$0.03 \\
Ni  I  & 6643.63  & 1.67 &   $-$2.30  &   75   &  3.69  &   $\mp$0.11   &  $\mp$0.01   &  $\mp$0.04 \\
Zn  I  & 4810.53  & 4.07 &   $-$0.13  &   30   &  2.19  &   $\pm$0.02   &  $\mp$0.12   &  $\mp$0.01 \\
Y  II  & 5200.41  & 0.99 &   $-$0.57  &  syn   &  0.00      &   $\pm$0.05   &  $\pm$0.10    &  $\pm$0.01 \\
Y  II  & 5205.73  & 1.03 &   $-$0.34  &  syn   &  0.00   &   $\pm$0.05   &  $\pm$0.10     &  $\pm$0.01 \\
Zr II  & 6127.48  & 0.15 &   $-$1.06  &  syn   & $<$0.50 &   $\pm$0.30   &  $\mp$0.30    &  $\pm$0.01 \\
Zr II  & 6134.59  & 0.00 &   $-$1.28  &  syn   & $<$0.50 &   $\pm$0.30   &  $\mp$0.30    &  $\pm$0.01 \\
Ba II  & 4934.08  & 0.00 &   $-$0.15  &  syn   & $-$1.20  &   $\pm$0.10   &  $\pm$0.10   &  $\mp$0.01 \\
Ba II  & 5853.68  & 0.60 &   $-$0.91  &  syn   & $-$0.50 &   $\pm$0.15   &  $\pm$0.20   &  $\mp$0.10 \\
Ba II  & 6141.73  & 0.70 &   $-$0.03  &  syn   & $-$0.90  &    $\pm$0.10   &  $\pm$0.10   &  $\mp$0.10 \\
Eu II  & 6645.09  & 1.38 &      0.12  &  syn   & $< -$1.50 &   $\pm$0.70   &  $\pm$0.50    &  $\pm$0.01 \\
\hline
\hline
 \end{tabular}
\label{tab:ews}
\end{table}
%
\end{appendix}
\end{document}